\documentclass{aastex}
\usepackage{emulateapj5,graphicx,epsfig,times}
\usepackage{onecolfloat}

\newcommand\kms{km~s$^{-1}$}
\newcommand\hkpc{$h^{-1}$~kpc}              %1/h kpc
\newcommand\hmpc{$h^{-1}$~Mpc}              %1/h Mpc
\newcommand\eg{{\it e.g.}}
\newcommand\rhl{$r_{\rm hl}$}

\slugcomment{{\sc Accepted for publication in The Astrophysical Journal: } \today}

\begin{document}
\def\head{

\title{Structural Evidence for Environment-Driven Transformation of the 
Blue Galaxies in Local Abell Clusters - A85, A496,
and A754}
%\footnote{Observations reported here were obtained at
%itt Peak National Observatory, National Optical Astronomy Observatories,
%hich is operated by the Association of Universities for Research in Astronomy,
%nc., under cooperative agreement with the National Science Foundation.}}

\author{
Daniel H. McIntosh\altaffilmark{2,3},
Hans-Walter Rix\altaffilmark{4}, and
Nelson Caldwell\altaffilmark{5}
}

\altaffiltext{2}{Steward Observatory, University of Arizona, 933 North Cherry Avenue, Tucson, AZ 85721}
\altaffiltext{3}{present address: University of Massachusetts, Lederle Graduate Research Tower, Amherst, MA 01003 -- dmac@hamerkop.astro.umass.edu}
\altaffiltext{4}{Max-Planck-Institut f\"{u}r Astronomie, Heidelberg, Germany -- rix@mpia-hd.mpg.de}
\altaffiltext{5}{Smithsonian Astrophysical Observatory, Cambridge, MA 02138 -- caldwell@cfa.harvard.edu}

\shorttitle{Transformation of Blue Cluster Galaxies}
\shortauthors{McIntosh, Rix, \& Caldwell}

\begin{abstract}
We present a detailed comparison of structural properties in
the rest-frame $V$-band
of cluster and field galaxies, selected and analyzed in the same manner,
to test the hypothesis
that much of the current cluster galaxy membership resulted from the
fairly rapid (1-2 gigayears) transformation of infalling, field spirals into
red, cluster early-types.  Specifically,
we have selected $\sim140$ galaxies from three nearby
Abell clusters (A85, A496 and A754) that have colors significantly bluer 
than the red sequence population, and compared them to $\sim80$
field galaxies with similar colors and 
luminosities from Jansen et al. (2000, ApJS, 126, 271).  
The comparison is based on the hypothesis that recent (1-4 gigayears)
cluster arrivals were originally blue and star-forming, then stopped
forming stars to dim and redden in a few gigayears.
For the comparison
we quantify galaxy internal structure and morphology
from two-dimensional bulge/disk decompositions using GIM2D.

We observe structural differences between blue galaxies in local
($z<0.06$) clusters, compared to field environments.  All cluster
galaxies have spectroscopic membership.
The majority of blue cluster members, presumably recent additions,
are physically smaller and fainter than their equally-colored
field counterparts.  At a matched size and luminosity, the
newer cluster arrivals are quantifiably smoother in appearance, yet their total
light is as disk-dominated as in normal field spirals.
Moreover, half of the blue cluster members appear to have blue cores or
globally blue color profiles, in contrast with field
spirals, which typically exhibit red inward color gradients.
Blue cores suggest enhanced nuclear star formation, possibly a starburst, while
uniformly blue profiles are consistent with an episode of fairly strong
global star formation in the past few gigayears.  Our previous work
(McIntosh et al. 2004, ApJ submitted) shows
that the blue membership of local clusters is a recently infalling population
that has yet to encounter the dense core.
In a Universe {\it without environmental dependent evolution
outside of cluster cores}, we would expect blue disk galaxies
inhabiting field and cluster regions to have similar
morphology, size, and color gradient distributions.
Our findings show conclusively that not only the abundance of red and blue
galaxies depends on environment, but also that fundamental structural
and morphological galaxy
properties do indeed reflect the environment in which the galaxy is found.  
Moreover, the data show that the transformation of accreted galaxies is
{\it not} confined to the dense cluster core.
The overall properties of bluer cluster members are best explained
by environment-driven transformation of accreted field spirals, and
our results suggest that the processes that govern
color and morphological evolution occur separately.

\end{abstract}

\keywords{
galaxies: clusters: general --- 
galaxies: clusters: individual (A85,A496,A754) --- galaxies: evolution ---
galaxies: fundamental parameters (colors, luminosities, morphologies, sizes) --- galaxies: structure}

} %end head
\twocolumn[\head]  %this *somehow* makes title,author,abstract shit be 1col.
                   %and body of text be 2col.

%-- 1 INTRO----------------------------------------------------------------
\section{Introduction}
If structure in the Universe forms hierarchically, the massive
galaxy clusters we observe locally should have been built through
the infall and accretion of matter ({\it i.e.} galaxies, groups, subclusters)
along large-scale filaments
\citep[][and references therein]{white78,white91,west95,kodama01c}
As new galaxies fall into regions of
greater density, they must interact with both the hot, inter-cluster
medium (ICM) and the increased numbers of nearby, cluster neighbors.
Given these interactions, it seems plausible that 
infalling populations undergoing ``normal'' star formation rates (SFR)
will be transformed into red lenticular (S0) galaxies
%ver short (few gigayears) timescales
\citep[hereafter the ``transformation scenario'',][]{gunn72,charlot94,abraham96,dressler97,vandokkum98,moore99,poggianti99}.
The transformation scenario has been invoked
to explain the well-known evolution of cluster galaxy color
\citep[the Butcher-Oemler effect, \eg][]{bo78a,bo84,rakos95,margoniner01,ellingson01}
and morphology
\citep{lavery88,couch94,dressler94a,dressler94b,ellis97,oemler97}.
This scenario is key to the idea of ``progenitor bias'' put forth by
\citet{vandokkum01}, which states that the progenitors of the youngest 
early-type galaxies in clusters
were morphologically transformed from later types, and thus would not
be considered as part of the early-type population of clusters at
larger redshifts.
Moreover, this environmental based scenario is compatible with
the observed correlations between increased density and cluster galaxy
morphology 
\citep[the ${\rm T}-\Sigma$ relation, \eg][]{melnick77,bo78b,dressler80,postman84,dressler97,hashimoto99}
and star formation (SF) properties 
\citep{balogh97,balogh98,hashimoto98,lewis02}.

Establishing whether the transformation scenario is occurring in low
redshift clusters
provides important confirmation for the continued hierarchical evolution
of these massive systems to the present epoch.  For example, finding cluster
members in the midst of this predicted evolution will secure the present-day
whereabouts of the blue galaxies that were more abundant in $z\sim0.3$
clusters (Butcher-Oemler effect), and may give us additional clues
to how and where galaxies evolve in dense environments. 
The literature is replete with evidence for the evolution and 
environmental dependence of cluster galaxy properties.  Yet, even with the huge
body of observational data on galaxy clusters in the local Universe,
there has been no
direct confirmation showing any fraction of their members in the midst of this
predicted galaxy transformation.  Therefore, this lack of confirmation
means either that such transformations are no longer ongoing in the present-day
Universe, or that the data and observational tests have been insufficient.  To
this end we have wide-field imaged three nearby Abell clusters with hundreds of 
spectroscopic member galaxies to look for evidence of the transformation
scenario.  In
\citet[][hereafter Paper~2]{mcintosh04} we established the existence of
significant fractions ($18-23\%$) of blue and moderately blue members with
kinematic and spatial properties as expected for a recently
infalling population \citep{diaferio01}.  Finding these populations of
late arrivals provided one crucial test of the validity of the transformation 
scenario.  Moreover, we found that many of the blue
members have disky visual morphologies, yet lack strong spiral
features, perhaps the result of morphological smoothing.
In this paper we use
quantitative measures of morphology and structure to determine whether the
appearances of the recent cluster arrivals differ from those of a control
sample of star-forming spirals from the lower density ``field''
environment.
Physical differences between cluster members and field galaxies with
similar star-forming blue colors will provide direct observational confirmation
for the morphological smoothing aspect of the transformation scenario.

Various physical mechanisms have been put forth to explain the transformation
scenario in clusters.
There is general agreement that the denser environment will have an
effect on the SF properties of new arrivals.  The cold
neutral gas supply may be removed by ram-pressure stripping during a galaxy's
virial motion through the hot ICM \citep{gunn72,solanes92,abadi99,quilis00},
or tidal stripping by either
the cluster potential or individual galaxies \citep{spitzer51,valluri90}.
Alternatively, the star-forming gas could be
rapidly consumed in a burst of SF triggered by ram-pressure effects
\citep{dressler83,fujita99}, tidal effects \citep{byrd90,hashimoto98,bekki99},
or galaxy merging \citep{barnes91,bekki98}.
Other authors have suggested
that the replenishment of star-forming gas is cut off instead, and that the
continuing formation of stars gradually ($>1$~gigayear) exhausts the remaining
supply of \ion{H}{1}
\citep{larson80,kodama01a,balogh00,balogh00b}.
Whether the SF fuel supply is removed, used up or cut off, the
result is the eventual end of SF and the passive fading and reddening of the
once blue galaxies as their hot, young stars die and their mean
stellar population age grows older.  This passive ``color evolution''
has a minimum timescale equal to the lifetimes of the short-lived luminous
stars ({\it i.e.} 1-2~gigayears for A-type stars), and longer if SF
is slowly diminished, rather than suddenly truncated. 

It stands to reason that
the cessation of SF in spirals will eventually produce an overall smoother
galaxy profile by reducing substructure ({\it e.g.} 
spiral arms, \ion{H}{2} knots),
yet this may take several gigayears or longer.  Some dynamical
simulations predict that galaxies falling into dense clusters will undergo
more rapid ($\sim1$~gigayear) changes in their physical
morphology due to encounters with other cluster members.  Among
processes that predict rapid spiral to S0 evolution are
``galaxy harassment'', the smoothing of galaxy appearances and stripping
of matter from 
multiple tidal encounters with other cluster members 
\citep{moore96,moore96b,moore98,moore99},
and unequal-mass spiral mergers \citep{bekki01,cretton01}.
Other processes, such as ``passive'' spiral formation from halo gas stripping
and subsequent ``starvation'', predict a much more gradual transition 
($\sim3$ gigayears) into S0s \citep{bekki02}.
Presently, it is unclear
whether color and physical morphology evolution occur simultaneously,
or if they are decoupled \citep{poggianti99,couch01,kodama01b}.

Typical Abell clusters have measured
velocity dispersions of $\sigma\leq1000$~\kms\ \citep[\eg][]{depropris02},
which correspond to roughly 1~Mpc per gigayear.
At this rate, recent arrivals would take several gigayears to make their
way from the cluster outskirts to the inner regions and, given the
timescales for the various proposed mechanisms,
much of the predicted rapid evolution would have already transpired.
Furthermore, the extreme environments within the inner half megaparsec of
clusters will destroy or seriously truncate disk galaxies \citep{moore98},
hence looking for recent arrivals and evidence of transformation towards
the centers of clusters is not profitable.
Unfortunately, until quite recently CCD imagers were
limited to small fields of view so that it was quite difficult to image
large angular regions of the sky.  Thus, the bulk of cluster imaging
work in the nearby Universe ($cz<15,000$~\kms) has concentrated on
galaxies belonging to the core.  For example, at the distance to
the well-studied Coma cluster, 2~Mpc in diameter is projected across 
two degrees of arc on the sky.  \citet{terlevich01} used overlapping
CCD frames to study a square-degree region of Coma that
still only provided a view of the inner $R=0.5$~Mpc.  They found no
evidence for a
population of recent arrivals in mid-transformation.  Yet,
observations of more distant ($z>0.3$) clusters have yielded 
cluster S0s with blue colors -- possible mid-evolution examples of
transforming blue spirals into red S0s
\citep{vandokkum98,rakos00,smail01}.

Therefore, our observing campaign has taken advantage of
recent wide-field imaging capabilities and large redshift surveys 
to explore galaxies at outer ($R>0.5$ Mpc) regions in three local Abell
clusters: A85 ($z=0.055$), A496 ($z=0.033$) and A754 ($z=0.055$).
Most cluster studies rely upon statistical corrections to
remove foreground and background galaxy contaminates; therefore,
cluster membership using such corrections becomes more uncertain at
larger cluster radii.  Using the large cluster galaxy redshift survey
of \citet{christlein03}, we have limited our study to only
galaxies confirmed as members with spectroscopic redshifts within
$\pm3\sigma$ of the mean cluster redshift.

To test the transformation scenario we look for structural differences
between blue galaxies residing in cluster and field environments
through a detailed analysis of their $V$-band structural properties from
two-dimensional bulge/disk (B/D) decompositions using GIM2D \citep{simard02}.
We assume that infalling galaxies
suffer SF truncation and evolve rapidly in color and luminosity; hence,
relative blueness gives us
an estimate of cluster accretion time and ``membership age''.  In Paper 2
we performed a detailed $U,V$ color-magnitude
relation (CMR) analysis for 637 galaxies belonging to the three clusters
in our sample.  The bluer members reside preferentially in the cluster
outskirts (typically $R>0.5$~\hmpc) and have kinematics suggesting a
non-virialized, infalling population \citep{diaferio01}.  
For this study we select the 143 members with $M_V\leq -17+5\log{h}$
and colors more than $2\sigma_{\rm CMR}$ bluer than
their cluster CMR, where $\sigma_{\rm CMR}$ is the measured CMR scatter
($<0.1$~mag) in $(U-V)$ color (Paper~2).  For comparison, we select 78
galaxies with similar colors and luminosities
from the Nearby Field Galaxy Survey (NFGS) of
\citet{jansen00}.  We degrade and
resample the field members to measure their properties as if they were
observed at $z=0.055$. 

In the transformation scenario we expect that the most recent cluster
arrivals will have bluish colors from recent (within $<2$ gigayears)
SF, disk-like morphologies reflective of an infalling spiral population,
and smoother appearances sometime after SF cessation.  In Paper 2 we find
most blue members in these clusters are classified visually as disk-dominated
systems with weak spiral features.  
SF and morphology evolution may be decoupled in cluster galaxies
\citep[e.g.][]{couch01}, and whether or not the newest (blue) members already
exhibit morphological smoothing provides a key motivation for this work.
Furthermore, processes like galaxy harassment
will reduce the relative disk substructure and size \citep{moore98}, 
and a variety of mechanisms predict centralized starbursts.
In our third paper
(hereafter Paper~3, D. H. McIntosh, H.-W. Rix, \& N. Caldwell, in preparation)
we establish $U$ and $V$-band structural differences between three 
color-magnitude (C-M) selected cluster populations.  The bluer membership found
typically at large cluster-centric distances consists of predominantly disky
systems with uniformly blue to blue centered profiles, lending additional
credence to their recent infall origin.  Of interest, we notice that the
blue members have little excess substructure relative to the redder,
presumably older members with smooth morphologies.  By looking for structural
differences between cluster and field blue galaxies, we can determine
whether recent cluster arrivals have undergone transformation compared
to their presumed progenitors, 
star-forming galaxies believed to be the infall source
in an hierarchical cosmology.  Here we concentrate our effort on
$V$-band properties given the higher signal-to-noise (S/N) and better
seeing characteristics of our cluster imaging in this passband.

In this paper we present evidence for environment-driven galaxy 
transformation
through a detailed comparison of cluster members and field galaxies
with similar luminosities and blue colors.  We briefly
summarize the cluster and field blue galaxy sample selections in \S2.
In \S3 we describe our two-dimensional B/D decompositions
to measure quantitative morphologies in both the cluster and field samples.
We include a detailed study of the limitations of this method
to yield useful morphological measurements from the cluster and field data.
In \S4 we analyze detailed comparisons for the observed properties of cluster 
members against field counterparts selected in the same manner.  We discuss
our results in light of morphological transformation processes in \S5.
And we give our conclusions in \S6.
Throughout this paper we use $h = H_0/(100$~km~s$^{-1}$~Mpc$^{-1})$, and we
assume a $\Lambda$-CDM (cold dark matter),
$\Omega_{\rm M}=0.3$ and $\Omega_{\rm \Lambda}=0.7$, flat ($\Omega_{\rm k}=0$)
cosmology. 

%-- 2 GALAXY DATA-----------------------------------------------------------
\section{Galaxy Data}
\subsection{Cluster Sample}
The cluster galaxy data come from wide-field 
(one square degree) $U$ and $V$ (Johnson system) imaging of three
local Abell clusters (A85, A496, and A754) using the NOAO Mosaic Image on
the Kitt Peak National
Observatory (KPNO) 0.9-meter Telescope.  Complete details of the
sample selection, observations, data reduction, photometric calibration,
and cluster membership catalog construction
are reported in Paper~2.  The clusters were selected primarily
due to the availability of a large spectroscopic redshift database
\citep{christlein03}
providing memberships in excess of 100 galaxies per cluster. 
The final sample contains a total of 637 spectroscopically confirmed
cluster member galaxies over a large range of absolute magnitudes
($-17 \lesssim M_V-5\log{h} \lesssim -23$).  In Paper 2 we provide a detailed assessment
of the completeness of our cluster membership selection (see Paper 2, Fig. 2).
Specifically, at $V\le18$ we have $U$-band photometry for
95\% (727/765) of our $V$-band extended source detections, and we find that 85\%
(615/727) of these have redshifts.
For each cluster, we give membership information in Table~1.

The full details of our galaxy photometry and CMR analysis are published
in Paper 2.  Briefly, we used maximum-likelihood fits to each
CMR to measure relative cluster galaxy color.
Compared to the population of old early-types that
define the mean CMR, galaxies with bluer integrated colors than expected
for a given luminosity are interpreted as having younger luminosity-weighted
mean ages \citep{vandokkum98,terlevich99}.  Given that clusters
form hierarchically through the accretion of star-forming, field galaxies
which then have their star production shut down once they become associated
with the parent cluster, it follows that more recent arrivals would
contain a younger stellar population and hence, have bluer colors
\citep{kodama01a,bicker02}.  Therefore,
the SF history (colors) of cluster galaxies appears to
depend on their time since arrival \citep{balogh00}.  

Based on the assumption that the color difference $\Delta (U-V)$ 
between the mean CMR and
a galaxy provides a rough division of arrival timescales, we separated
the cluster galaxies into three color-based populations to identify
newer cluster members (see Paper 2 for details):
\begin{enumerate}
\item Red sequence galaxies (RSGs) with colors redward of
$-2 \sigma_{\rm CMR}$.
\item Intermediately blue galaxies (IBGs) with
$-2 \sigma_{\rm CMR} > \Delta (U-V) > -0.425$~mag.
\item Very blue galaxies (VBGs) with $\Delta (U-V) \leq -0.425$~mag.
\end{enumerate}
The VBG cutoff corresponds to the \citet{bo84} criteria
of $\Delta (B-V)=-0.2$~mag, and represents spiral-like integrated colors. 
We defined a second, somewhat blue population (IBGs) with the aim of locating
galaxies belonging to clusters for an intermediate time between the recently
arrived VBGs and the majority population of long resident red galaxies.
We selected $-2 \sigma_{\rm CMR}$ as the dividing line between the red cluster
galaxies and the IBGs because it provides a good match to the blue envelope
of the RSGs and it is roughly midway between the default CMR and 
the \citet{bo84} criteria.  In Paper~2 we established that
the blue (IBG$+$VBG) members make up $18-23\%$ of each cluster population more 
luminous than $0.1L^{\star}$.\footnote{We assume
$M^{\star}_V = -20.6 + 5\log{h}$ for red-sequence galaxies from
$M^{\star}_B = -19.7 + 5\log{h}$ \citep{binney98} and a mean E/S0 galaxy color
of $(B-V)=0.90$~mag \citep{fukugita95}.}  
Moreover, we found that the spatial and
kinematic properties of blue cluster galaxies, taken together, are distinct
from RSGs confirming the late arrival nature of the blue members.

We give the number of blue members in each cluster in Table~1.  To look for
evidence of transformation in more recent cluster arrivals, we focus in
this paper on the
observed properties of the blue cluster populations in relation to field
galaxies selected with similar colors and luminosity.  For this reason
we limit the cluster IBG and VBG samples to galaxies brighter than
$M_V=-17+5\log{h}$ to match the field sample selection (next section).  This
selection results in 143 blue members, the ``cluster sample'', which we
use throughout this paper.  We discuss the
structural properties of RSG members in Paper~3.

\subsection{Field Sample}
\label{fieldsel}
To construct a useful field galaxy comparison sample we select
galaxies from
the Nearby Field Galaxy Survey (NFGS) of \citet{jansen00}.  The NFGS was
selected from the first CfA redshift catalog of \citet{huchra83}. 
The actual Johnson $B$-band and Cousins $R$-band CCD
images for 195 of 198 NFGS galaxies were made available by Rolf Jansen.
The NFGS provides a representative sample of local
($\sim95\%$ have $cz<10,000$~\kms, with a median of 3000~\kms), 
field galaxies spanning
the full range of morphological types along the Hubble sequence.  With
similar luminosities and rest-frame $(U-V)$ colors as our cluster
galaxy data, this sample makes an ideal data base for comparing galaxies
residing in different environments.  We point out that Jansen et al.
consider ``field'' to cover a wide range of environments with no
specific local density criteria, except for purposefully leaving out
Virgo cluster galaxies.  Although the NFGS is not a purely low density
sample, we are satisfied that it represents non-cluster
galaxies.

Prior to applying a C-M cut,
we discard all 27 NFGS galaxies
with irregular or peculiar morphological classification 
\citep[T-type$\geq9$,][]{jansen00}.  Irregular/peculiar surface 
brightness profiles often have multiple flux peaks and no clearly defined
center resulting in poor or failed profile fits.  We will compare
cluster and field member structural properties derived from two-dimensional
surface photometry fitting using GIM2D (described in \S3).
In Paper~3 we found
that the majority of cluster members were reasonably well fit by
a combined bulge and disk profile because these galaxies
have ``normal'' (not irregular) light distributions.  Although the
removal of irregulars could skew the overall field sample structural
morphology distribution towards lower values,
we will show that a strong difference in the amount of
residual substructure observed in cluster and field systems remains.

An additional 30 NFGS
galaxies are left out due to foreground stars that are superimposed on
or near the galaxy's disk.  Line-of-sight foreground stellar contamination is
difficult to model and, if not properly accounted for, will result
in poor profile fits using GIM2D.  We remove 14 field
galaxies with Local Group standard of rest velocities 
$cz_{\rm LG}<1300$~\kms.  To make objective comparisons to our cluster
data, we must artificially redshift and reobserve each field galaxy image
to simulate its appearance at the distance to our clusters.  The few NFGS
objects closer than this velocity criteria are too poorly sampled
following this procedure and, thus, are of no use in our comparative
analysis.  Finally, we leave out 3 NFGS galaxies with irregular or
Seyfert nuclei which will be difficult to model, 
and 3 with $M_V>-17+5\log{h}$~mag 
corresponding to the faintest cluster galaxies.  We note that 14 of the 27
irregular/peculiar (T$\geq9$) galaxies cut from the field sample have
$M_V>-17+5\log{h}$~mag.

\begin{figure}[hp]
\center{\includegraphics[scale=0.65, angle=0]{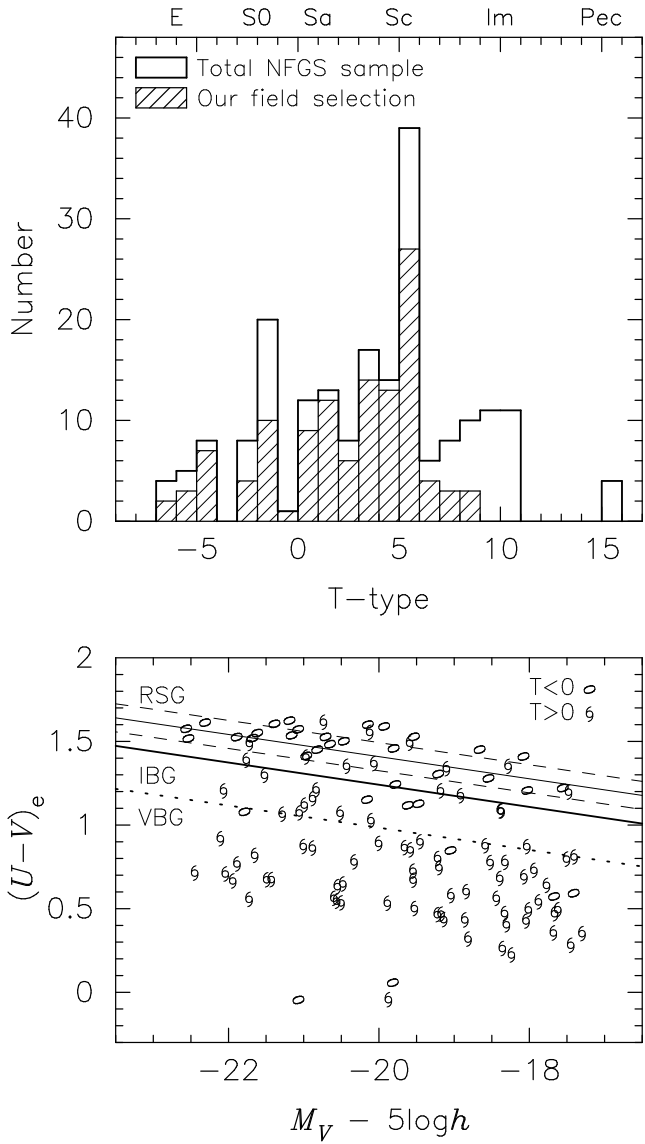}}
\caption
{General properties of our field galaxy comparison sample 
selected from the NFGS \citep{jansen00}.  {\it Top panel:}
Morphological T-type distribution for the 117 galaxies selected and analyzed
here ({\it hashed bins}) relative to the total NFGS
distribution ({\it solid line}).  The majority of blue NFGS galaxies
not included are either classified in \citet{jansen00} as
irregular (T-type$\geq9$), have Local Group standard of rest velocities
$cz_{\rm LG}<1300$~\kms, or have foreground stellar contamination. 
{\it Bottom panel:} Rest-frame C-M diagram for the 117 field
galaxies we have selected.  The $(U-V)$ colors are estimated (see text for
details).  Early-type (${\rm T}\leq0$) and late-type
(${\rm T}>0$) objects are distinguished with representative symbols.  We
plot the mean CMR for cluster A754 ({\it solid line}), 
with $\pm1\sigma_{\rm CMR}$ scatter ({\it dashed lines}), used in
Paper~2 to define C-M based galaxy population bins.  The
{\it bold solid line}, at $2\sigma_{\rm CMR}$ blueward of the mean CMR,
represents the red/blue galaxy cut.  There are 36 red sequence galaxies
(RSGs) and 78 blue (IBG and VBG) galaxies.
\label{FldSamp} }
\end{figure}

Our selection leaves 117 field galaxies split into 35 early-types
(${\rm T}\leq0$, E-S0/a) and 82 late Hubble types
(${\rm T}>0$, Sa-Sdm), based on qualitative classifications given
in \citet{jansen00}.  We make no additional morphological cuts because
our C-M criteria for blue galaxies does not necessarily exclude early-types
such as E/S0s.
We plot the T-type distribution for our field galaxy
selection compared to the total NFGS distribution in
the top panel of Figure~\ref{FldSamp}.  For early and late types our selection
provides a similar distribution as the total NFGS over $-7\leq T\leq7$; i.e.,
a simple K-S test finds a probability of 49.4\% that the two distributions
are different over this range of T-types.

Next, we apply our $U,V$ C-M criteria to select a blue field galaxy sample.
One drawback to using the \citet{jansen00} sample to compare
to our cluster data is that it has been imaged in $UBR$;
therefore, it is necessary to transform from $(U-B)$ to $(U-V)$ colors.
First, using the redshift $z_i$ of each galaxy, we convert 
the \citet{jansen00} $B$-band apparent magnitude to an absolute
$M_B$ magnitude using the k-correction from \citet{poggianti97}
and a cosmological distance modulus \citep{hogg00}, assuming 
$\Omega_{\rm M}=0.3$, $\Omega_{\rm \Lambda}=0.7$, and $h=1$.  The given apparent
magnitudes were corrected for Galactic extinction using the \citet{burstein82}
\ion{H}{1} maps.  Next we estimate
the $M_V$ of each field galaxy using the following color relation:
\begin{equation}
(B-V)_e \approx 0.65(B-R)_e - 0.06 ,
\label{deJongBV}
\end{equation}
where the effective $(B-R)_e$ colors (measured within elliptical half-light
radii) are given in \citet{jansen00}.  This color relation is based on
Bruzual and Charlot single
burst models over a full range of age and metallicities, and is
consistent with the
mean integrated colors of all spirals in \citet{dejong96c}.  This color
translation has an estimated uncertainty of $<0.05$~mag.  Last the
field sample $(U-V)_e$ colors are estimated from $(U-B)_e$ provided
in the NFGS.  We show the estimated $U,V$ C-M diagram for the field
sample of 117 galaxies in the bottom panel of Figure~\ref{FldSamp}.  
With the IBG and VBG population 
divisions defined using the mean CMR of A754, we select the
final subset of 78 blue galaxies -- the ``field sample''.  
These galaxies span an absolute
magnitude range of $-22.5<M^{\star}_V-5\log{h}<-17.3$ and a
redshift range of $0.05<z<0.041$.
The field sample has an excess of bright
($L>L^{\star}$) galaxies and a deficit
of fainter systems relative to the cluster sample.  In \S4.4.3 we show that
the difference between bright cluster and field VBG distributions is
representative of these two environments, 
while the deficiency of faint blue field
galaxies is partially due to our culling of irregular galaxies.  Our
findings are not affected by the difference in numbers of faint blue
galaxies.

We want to look for differences in amounts of
residual substructure, left over after the best 
GIM2D fits to our $V$-band cluster images and to the field sample images.
In spiral galaxies the strength of residual and
asymmetric features (spiral arms, \ion{H}{2} knots, etc.) is wavelength
dependent because such features are predominantly due to recent and localized
SF, which is more prominent in blue light.  If recent cluster arrivals (bluer
members) are being transformed, 
we expect to measure significantly less substructure
in these galaxies, relative to field counterparts of similar color
and luminosity.  To compensate for the lack of $V$-band NFGS imaging,
we will estimate $V$ field galaxy parameters by averaging the results 
from fits to the $B$ and the $R$ passband data straddling the $V$-band.
We opt to concentrate our analysis on $V$-band properties, rather than
$U$, owing to the improved S/N and seeing characteristics
of our cluster imaging in this passband.

%fig1

%-- 3 2D B-D DECOMPOSITIONS-----------------------------------------------
\section{Two-dimensional Bulge$+$Disk Decompositions}
For our analysis we use GIM2D (Galaxy IMage 2D)\footnote{Available at http://www.hia.nrc.ca/STAFF/lsd/gim2d/}
v2.2.1 to perform multi-component model fits to the
two-dimensional (2D) surface brightness distribution of each galaxy in
the cluster and field samples.  Hereafter we will refer to the 2D galaxy 
profile fitting and parameterization as bulge/disk (B/D) decomposition.
A full description of GIM2D is given in \citet{simard02}.
We select GIM2D for the following reasons: (i) it
accounts for the effects of seeing by convolving 
the best-fit model with an input point-spread-function (PSF);
(ii) it provides a variety of model profiles including the traditional
\citet{devaucouleurs48} $r^{1/4}$ bulge, plus exponential disk; and
(iii) it is well-tested as evident in a growing body of published galaxy
structure work using GIM2D
\citep[\eg][]{simard99,im01,im02,tran01,balogh02a,balogh02b,simard02}.  Furthermore,
a true 2D fitting routine is advantageous over traditional one-dimensional
methods because it quantifies 
non-axisymmetric information contained in galaxy images, such as
residual flux from spiral arms \citep{dejong96a}.

\subsection{Structure and Morphology Measurements}

We have used GIM2D to fit an $r^{1/4}$ bulge and an exponential disk
to the surface brightness profiles of our sample of
637 known members from $U,V$ imaging of clusters A85, A496, and A754.
For this analysis we concentrate on the GIM2D fitting to our $V$-band imaging.
Our B/D decomposition procedure and a detailed analysis of the internal
structure of red and blue cluster members are given in Paper~3.  
Briefly, GIM2D fits each sky-subtracted galaxy (``postage stamp'') image with a
PSF-convolved model.  We use {\sc daophot} to construct a variable, high
S/N PSF model for each cluster using hundreds of bright but
unsaturated stars from its $V$-band mosaic image.  This model reproduces
the PSF ``shape'' for each galaxy.  Through detailed testing
we find that this PSF is sufficient for this galaxy fitting (see Paper 3).
During fitting GIM2D holds the background to a constant zero value.
Here we add B/D decompositions of the 78 blue galaxies selected from the NFGS.
As stated in \S2.2, we estimate galaxy properties in the $V$-band from
fits to the $B$ and $R$ passband images.  For the field sample we employ
the same fitting procedure as for our cluster imaging, with one exception --
for each NFGS galaxy image we use a 2D Gaussian matched to the
seeing size for PSF convolution.  The seeing FWHM is given in the image
header of each NFGS image and represents an average of several nearby stars.
Many of the NFGS images lack stars of
sufficient S/N and sampling to construct a useful PSF; therefore, we elect
to always use a Gaussian for consistency.
We compare galaxy parameters derived from fits to NFGS images using Gaussian
PSFs and neighboring bright stars (when they are available), and we find
little ($\sim5\%$) difference.

\begin{figure}
\includegraphics[scale=0.5, angle=0]{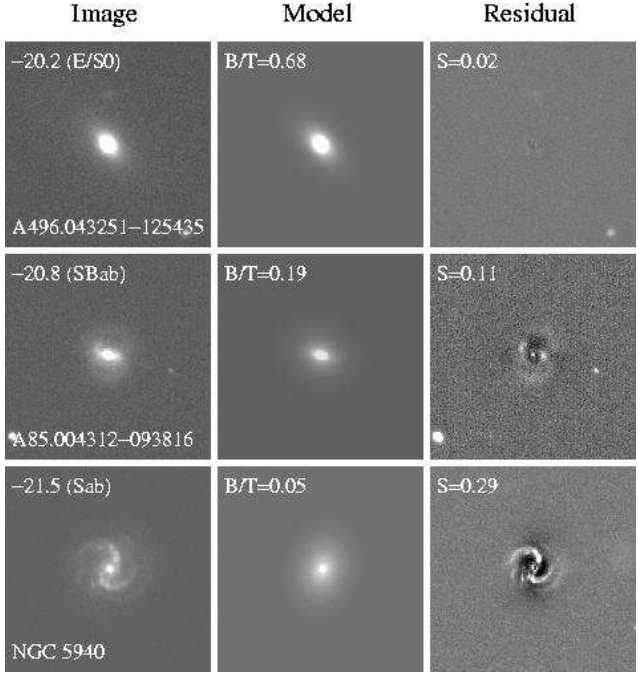}
\caption
{Example images illustrating quantitative morphology
measurements from GIM2D fits to three galaxy images spanning a range of
$B/T$ and $S$ values.  {\it Left panels:} object image with rest-frame $M_V$,
qualitative morphology, and galaxy designation printed.  The top and middle
rows are $V$-band thumbnails
from the cluster sample with identifications based on J2000.0 celestial
coordinates; the last row is NGC 5940 ($z=0.034$) from the NFGS in the $B$-band.
{\it Middle panels:} corresponding best-fit
GIM2D bulge$+$disk model image with bulge fraction $B/T$ value.  
{\it Right panels:} residual 
(object - model) flux image with substructure $S$ value.  The dynamic range
of the residual images have been increased to make faint
structural features more apparent.  $S$ quantifies the
amount of residual substructure from small values ($S\sim0$) for smooth galaxies
like E/S0s, to large measurements ($S>0.1$) for galaxies such as spirals.
\label{gim2d} }
\end{figure}

For each
galaxy, GIM2D produces model and residual (model-subtracted) images
(see Figure~\ref{gim2d}), along with a set of basic structural parameters,
including their internal confidence limits.  Measurements
of internal structure include size estimates
such as the bulge effective radius $r_e$, disk scale length $h_0$, overall
profile half-light radius $r_{\rm hl}$; bulge ellipticity $\epsilon$;
disk inclination $i$; and the position angles of each component ($\phi_{\rm B}$,
$\phi_{\rm D}$).  The $1\sigma$ internal errors for size measurements
are typically quite small with a mean of $\sim5\%$ for the cluster galaxy
fits.
In addition, GIM2D provides quantitative and repeatable 
morphological classification of cluster members through the bulge-to-total
ratio $B/T$ and residual flux measures (asymmetric and total) of galaxy
substructure $S$.  This two-parameter set lends itself to
our search for signs of recent evolution in cluster galaxies.  $B/T$
provides a measure of morphology ({\it e.g.} is the galaxy spheroidal or 
disk-like).  $S$ gives
a measure of how much, or how little, substructure ({\it e.g.} spiral features)
is present within a galaxy.  Our B/D decompositions have two limitations:
(i) we assume a constant $n=4$ bulge model profile; and (ii) we have no bar 
component.  We discuss these limitations in detail in Paper~3.

We use the quantitative morphology measurements ($B/T$ and $S$)
to explore possible differences between
cluster and field galaxies with similar photometric properties.
The use of quantitative morphological classification is
advantageous over visual (qualitative) classification because it is
reproducible and it is objective.
\citet{im02} show that the GIM2D $B/T$ and $S$
parameters are successful at selecting
E/S0s, without substantial contamination from later-type galaxies, from
a sample of local field galaxies.  $B/T$ measures how centrally concentrated
the galaxy surface brightness profile is by quantifying the relative
contributions of the bulge and disk model components to fitting the overall
galaxian light profile.  The residual substructure parameter $S$ measures
how discrepant a galaxy's surface brightness is from the simple 
bulge$+$disk (B$+$D) model.
Thus, E/S0s will typically have larger
$B/T$ measurements and $S$ near zero, whereas spirals should have low bulge
light fractions and greater measurable substructure.

Following \citet{im02},
we quantify the substructure contained within $r = 2r_{\rm hl}$ and define it as
\begin{equation}
S = R_T + R_A .
\end{equation}
From \citet{simard02} we have
\begin{equation}
R_T = \frac{\Sigma \onehalf | R_{ij} + R_{ij}^{180} |}{\Sigma I_{ij}} - \frac{\Sigma \onehalf | B_{ij} + B_{ij}^{180} |}{\Sigma I_{ij}} ,
\end{equation}
and
\begin{equation}
R_A = \frac{\Sigma \onehalf | R_{ij} - R_{ij}^{180} |}{\Sigma I_{ij}} - \frac{\Sigma \onehalf | B_{ij} - B_{ij}^{180} |}{\Sigma I_{ij}} .
\end{equation}
$R_{ij}$ is the residual image flux at pixel ($i,j$)
and $R_{ij}^{180}$ is the
($i,j$) pixel flux in the residual image rotated by 180 degrees.
Similar flux measures for the background noise, $B_{ij}$ and $B_{ij}^{180}$,
are calculated over an area comparable in size to the model and drawn
randomly from the full set of all background pixels in each postage stamp
image.  $I_{ij}$ is
the galaxy image flux at pixel ($i,j$).  Morphological
features such as spiral arms, bars, \ion{H}{2} regions, active nuclei or
dust lanes produce measurable residuals,
hence, $S$ provides a measure of the {\it total} substructure (deviations
from a smooth model profile) in a galaxy.  Note that $S$ is
not a measure of the total residual flux.  $R_T$ represents the total
summed {\it positive and negative}
residuals, while $R_A$ quantifies the absolute value of the
asymmetric residuals (\eg\ asymmetric spiral arms or \ion{H}{2} regions).
Therefore, a galaxy that exhibits
symmetric positive and negative residuals of equal flux would result in an
$S>0$ specifying correctly its inherent substructure.

To illustrate GIM2D's ability to quantify morphologies,
we show images of three example galaxies, along with their corresponding
best-fit models and residual frames in Figure~\ref{gim2d}. 
The galaxies span a range of $B/T$ and $S$ measurements.  For each galaxy
we include the rest-frame $M_V$, qualitative morphology, $B/T$, $S$,
and identification.  Two galaxies are $V$-band images from our cluster
data, the third (NGC 5940, $z=0.034$) is $B$-band data from the NFGS.
GIM2D does an excellent job 
reproducing the general appearance of each galaxy's bulge and disk 
as depicted by the 2D model images.  Moreover, the amplitude of residuals
as measured by $S$ provide a reasonable method for quantifying the
amount of substructure observed, from little residual flux ($S\sim0$ for an
E/S0 fit) to large $S$ values due to strong spiral features.

\begin{figure}[hp]
\center{\includegraphics[scale=0.65, angle=0]{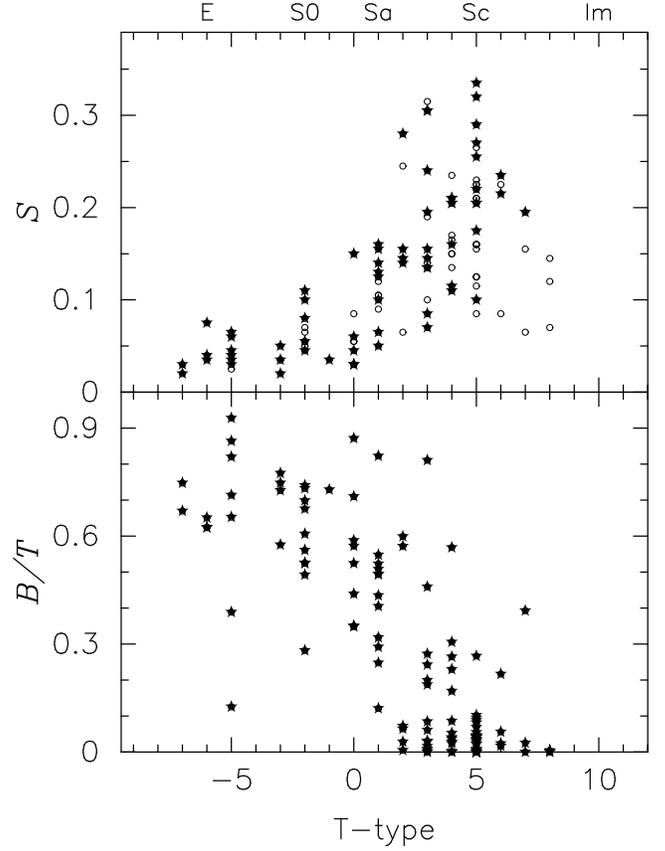}}
\caption
{Mean $B/T$ and $S$ quantitative morphology
measurements as function of T-type for the local field galaxy comparison
sample.  The 117 field galaxies selected prior to a red color cut are plotted
here.  Each parameter is given by the average of two values from GIM2D fits 
to $B$ and $R$-band images.
For the numerical T-types the 
corresponding Hubble types are provided along the top axis.  Galaxies that
are brighter than the reliability magnitude cut ({\it bottom:} 117/117 
$M_V\leq-17.0+5\log{h}$
for $B/T$; {\it top:} 67/117 $M_V\leq-19.5+5\log{h}$ for $S$)
are plotted as {\it filled stars}; those fainter are {\it open circles}.
Average errors in $B/T$ range from 25\% for $B/T<0.15$, to
10\% for $B/T>0.30$; and $S$ has estimated uncertainties of $10-20\%$.
\label{TvsQM} }
\end{figure}

In Figure~\ref{TvsQM} we plot the mean quantitative morphologies 
from our fits to the $B$ and $R$-band field galaxy images
as a function of qualitative (``by eye'') classifications given in
\citet{jansen00}.
The bottom panel shows a trend between $B/T$ and T-type in the sense that
early-type galaxies are bulge-dominated while the fraction of bulge
light diminishes to $B/T\sim0$ for late-type spirals.  This rough
correlation has been documented in other studies 
\citep{dejong96b,donofrio01,graham01}\footnote{In these studies the authors
use $B/D$ the bulge-to-disk luminosity ratio which is directly related to 
the bulge fraction by $B/T = \slantfrac{B/D}{B/D +1}$.}; however,
this trend has large scatter and does not provide a good indicator of
Hubble type \citep{dejong96b,lilly98,im01,im02}.  We find an $S$-T correlation
(top panel) with similarly large scatter.  Recall that $S$ is a measure
of total deviations from a smooth model profile and, hence, quantifies the
amount of substructure due to a variety of processes.  In disk galaxies (${\rm T}>0$)
the dominate source of substructure is spiral features which represent SF
process, thus, the $S$-T correlation follows
the SFR dependence on Hubble type
\citep{kennicutt94,kennicutt98,james03}.

\begin{figure}[hp]
\center{\includegraphics[scale=0.6, angle=0]{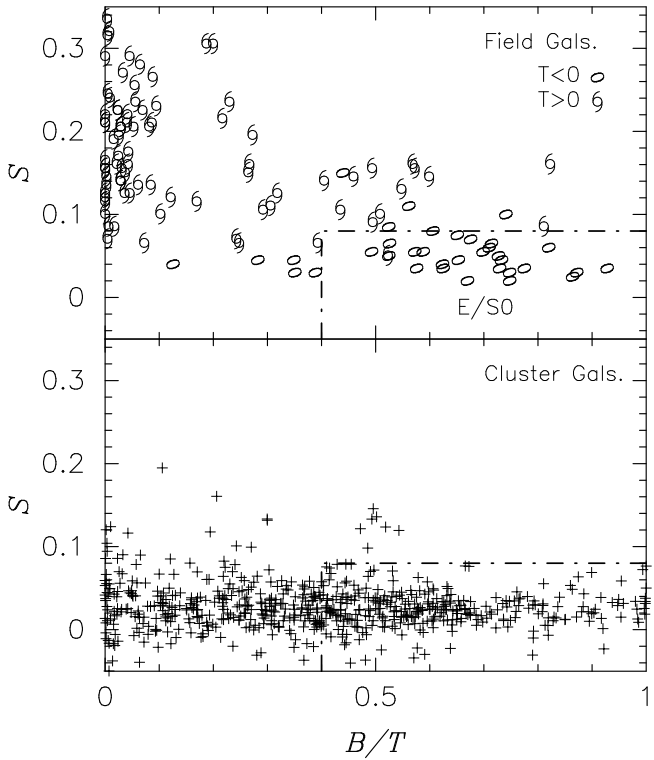}}
\caption
{$B/T$ versus $S$ quantitative morphology plane for the field ({\it top})
and cluster ({\it bottom}) samples.  The cluster results (637 galaxies) are
from GIM2D fits to the $V$-band imaging (Paper~3),
and the field data (117 galaxies)
are average morphology parameters from $B$ and $R$-band fits.  Mean errors
are 25\% ($B/T<0.15$) and 10\% ($B/T>0.30$), and we estimate an 
error in $S$ of $10-20\%$.
The region bounded by the $B/T\geq0.4$ and
$S\leq0.08$ ({\it dot-dashed lines}) is known to correlate well with E/S0
galaxies \citep{im02}.  The field galaxies are given symbols representing
early (${\rm T}\leq0$) and late (${\rm T}>0$) types as in Figure~\ref{FldSamp}.
For ease of viewing, all
cluster galaxy $S$ values have been randomly shifted by
$\epsilon \in (-0.01\leq \epsilon \leq 0.01)$.
\label{BTSplane} }
\end{figure}

In Figure~\ref{BTSplane} we show the cluster and field samples plotted in the 
quantitative morphology plane.  The cluster results are from GIM2D fits
to the higher S/N
$V$-band data for the complete cluster membership (including RSGs) from Paper~3.
The mean uncertainties in $B/T$ depend somewhat on the value of this
parameter: $\sigma_{B/T}\sim25\%$ ($B/T<0.15$), 
$\sigma_{B/T}\sim15\%$ ($0.15<B/T<0.30$), and
$\sigma_{B/T}\sim10\%$ ($B/T>0.30$).  GIM2D does not estimate an internal
confidence interval for the residuals $R_A$ and $R_T$, thus, we estimate
an average error of $10-20\%$ in our $S$ measurement from multiple fits
to a subset of galaxies.
The region of
$B/T$-$S$ space occupied by E/S0 types in \citet{im02} is shown for
both samples. 
\citet{im02} showed that $B/T\geq0.4$ and
$S\leq0.08$ provide a good method for discriminating E/S0s from other
morphological types in lower S/N data.  
We note a handful of $S<0$ cluster galaxies exist in Figure~\ref{BTSplane}.
Such measurements result from larger background values relative to the flux
residuals (see equations 3 \& 4).  In our subsequent analysis we consider
these galaxies to be morphologically smooth and fix their total 
substructure to $S=0$.  Finally,
there is significantly more observed substructure
in the field galaxies than the cluster members.  Nearly all 
disk-dominated ($B/T\leq0.3$) cluster galaxies have $S<0.1$, in stark
contrast to the disky later types in the field.
This finding is consistent with the low spiral fractions ($<20\%$)
found in local, rich clusters \citep[\eg][]{oemler74,bo78b}.  Yet,
the difference we observe may be merely due to the different observational
characteristics of these two samples.  This gross comparison
underlines the necessity to test what effect
our observations play on the differences in observed substructure.

%fig2

%fig3

%fig4

\subsection{Limitations of Morphologies Derived From the Cluster Data}
It is important to understand the effects that observational resolution
and S/N have on quantitative morphology measurements,
especially those that rely on residual fluxes from a smooth best-fit model 
profile.
The telescope aperture, pixel scale, seeing, and object distance 
all determine the effective
resolution and minimum surface brightness detection limit for a set of
galaxy observations.  Therefore, faint features such as spiral arms in
a flocculent disk galaxy may be measurable locally,
yet not seen in a similar galaxy at a somewhat larger distance when using
the same telescope and detector system under the same observing conditions.
The strong difference in $S$ values derived from our field and
cluster galaxy fitting (see Figure~\ref{BTSplane}) may be simply because
cluster members are observed through a smaller (1-meter)
telescope, and at distances twice as
far on average as the field sample.  Therefore, we must quantify the
limitations of our quantitative morphology measurements to establish the
significance of any evidence for recent evolution among cluster members.
To this end, we artificially place
the local field sample galaxies at a redshift corresponding to our most
distant cluster (A85, $cz=16,607$~\kms), simulate reobservation through
our 0.9-meter/Mosaic configuration, and repeat our B/D decompositions
on the re-imaged data.  Artificially
redshifting and reobserving galaxy data is a straightforward procedure
outlined by various authors \citep[\eg][]{tran01}; however, to produce our 
artificially redshifted field images requires an
extra level of complexity because we are simulating reobservation through
a different telescope/detector configuration than originally used.  We describe
our method in Appendix~A. 

\begin{figure*}
\includegraphics[scale=0.85, angle=0]{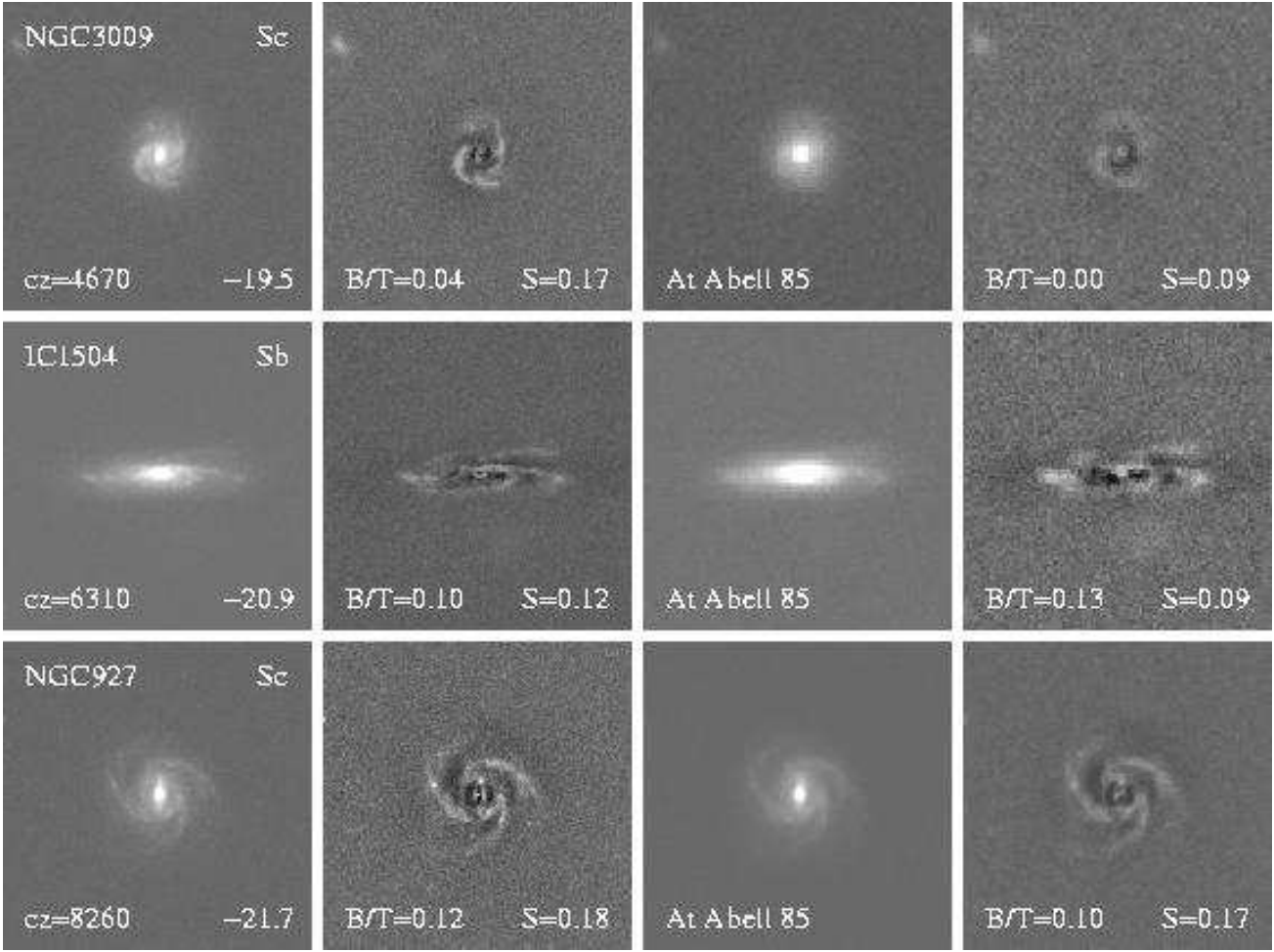}
\caption{
Images of three NFGS \citep{jansen00} spirals, spanning a representative
range in $M_V$ and redshift.
The panels for each galaxy from left to right show the original
observed-frame $R$-band image,
the residual spiral features after subtracting the best-fit B$+$D GIM2D
model, the galaxy's appearance at the distance of A85 ($z=0.055$) simulating
our $V$-band observational characteristics (0.9-meter/Mosaic system), 
and the GIM2D residual from fits
to the artificially redshifted image.  The galaxy name, Hubble type, $cz$,
and estimated $M_V-5\log{h}$ are printed in the original images ({\it left}).
The $B/T$ and $S$ parameters are given for the original and
redshifted residual images.  The sequence of galaxy images
illustrates how we determine the limitations of $B/T$ and $S$ to
distinguish morphological information for
our cluster observations.
\label{JansenExs} }
\end{figure*}

We determine the limitations on the parameters $B/T$ and $S$ for
distinguishing morphological information in our cluster observations by
comparing these morphology measurements from B/D decompositions for
field galaxies in their observed frame
and artificially placed at distances corresponding to cluster A85.  At
this distance we are illustrating the maximum degradation in
the $B/T$ and $S$ measurements from our cluster data.
The effect that our redshifting and reobserving simulations have on
the measured quantitative morphologies for
three spirals from our field galaxy sample are shown in Figure~\ref{JansenExs}.
For each field spiral we show its original NFGS observed-frame image, the
corresponding GIM2D residual flux image, its re-imaged appearance
through our 0.9-meter/Mosaic $V$-band system at the distance of A85, and
the GIM2D model subtracted residual for the artificially redshifted image.
Not surprising, 
the features of galaxies reobserved at further distances are washed
out compared to their original image counterparts.  In these three
cases, $S$ is more strongly affected (reduced) by our simulations 
than $B/T$, and the difference between original and reobserved
morphology measurements appears most dependent on object luminosity, at
a given distance.
We note that in all
three cases, the spiral substructure, after artificially redshifting
to the distance of A85 and reobserving through our 0.9-meter/Mosaic
system, is still conspicuous ($S\geq0.09$).  Therefore, while it is true
that our cluster observations are somewhat limited for quantifying the full
extent of potential substructure in spiral galaxies, the differences in the
distribution of $S$ between disk galaxies in the field and in the cluster 
(see Figure~\ref{BTSplane})
are {\it not} entirely due to the limitations of our observations.

To establish a brightness limit for achieving reliable measures of 
$B/T$ and $S$, we study the luminosity dependence at a given distance
of how quantitative morphologies measured from original field galaxy
images using GIM2D are diminished after redshifting and reobservation.

\subsubsection{Reliable $S$ Measurements}
\label{Stest}
Here we quantify how well our observations can detect the existence of
spiral substructure in galaxies at the distance to cluster A85.
%ere we quantify our ability to detect the existence
%f spiral substructure in galaxies from our cluster observations.
First we compare distributions of $S$ measured from blue-selected 
(IBG$+$VBG) field galaxies in observed and
redshifted frames, subdivided into four $M_V$ bins of roughly equal numbers
(see Figure~\ref{QMdistr}, top).  Recall that each galaxy $S$ value is
the average from GIM2D fits to $B$ and $R$-band data.
For the field IBG and VBG galaxies brighter
than $M_V=-18+5\log{h}$, the $S$ distributions from fits to the original data
appear qualitatively similar from one brightness bin to the next --
a broad distribution spanning $0.0<S<0.3$ with mean $S\sim0.15$.
A comparison to the
re-imaged $S$ distributions shows an abrupt change for $M_V>-19.5+5\log{h}$ such
that fainter galaxies moved to A85 appear to have much less
observed substructure ($S<0.1$) than seen in the original observations.
A K-S test finds the rest and redshifted $S$ distributions are $>99.9\%$
different in the two faintest bins.

In addition, we calculate the difference $\Delta S = S - S^{\prime}$ 
between substructure
measures from fits to the original ($S$) and artificially redshifted 
($S^{\prime}$) images; this difference represents how much spiral
morphology information
is ``lost'' when reobserving galaxies with known morphological features
at a greater distance.  We plot $\Delta S$ for each field galaxy in our
sample against $M_V$ in the top panel of
Figure~\ref{QMdel}.  We calculate the average change in substructure 
$<\Delta S>_M$ per $\Delta M_V = 1.0$ magnitude bin for only the IBG$+$VBG 
galaxies over the full range of field galaxy
luminosities.  There is a clear luminosity dependence on $\Delta S$, such
that $\Delta S$ increases in amplitude and spread for intrinsically fainter 
galaxies.
Nevertheless, $<\Delta S>_M$ does not deviate from zero by more than 0.05
until the $-20<M_V-5\log{h}<-19$ bin, corresponding to the same brightness where
we see gross differences in the overall $S$ distributions of original
and redshifted images (Figure~\ref{QMdistr}).

With this analysis we demonstrate that observing
a galaxy at a further distance reduces its S/N and increases
the physical size each pixel samples (pixel smoothing) -- {\it i.e.}
spiral galaxies with intrinsic substructure appear smoother.  These effects
lead to an underestimated measure of $S$ from GIM2D fitting.
If we assume that the brightness of
features, such as spiral arms or \ion{H}{2} regions, correlates with a
galaxy's overall luminosity, then we may determine the
average luminosity at a given distance at 
which this effective smoothing inhibits our ability
to quantify the existence, and absence, of substructure in galaxies
from our cluster $V$-band observations.  Provided this assumption holds and
based on our above analysis, we select $M_V=-19.5+5\log{h}$ to be the
brightness limit for measuring reliable
substructure in our cluster sample.  In Figure~\ref{TvsQM} we
distinguish field galaxies brighter and fainter than this limit using
separate symbols; however,
considering only reliable quantitative morphology measures does not reduce
the large scatter found between $S$ and Hubble type.

\subsubsection{Reliable $B/T$ Measurements}
\begin{figure}[hp]
\includegraphics[scale=0.5, angle=0]{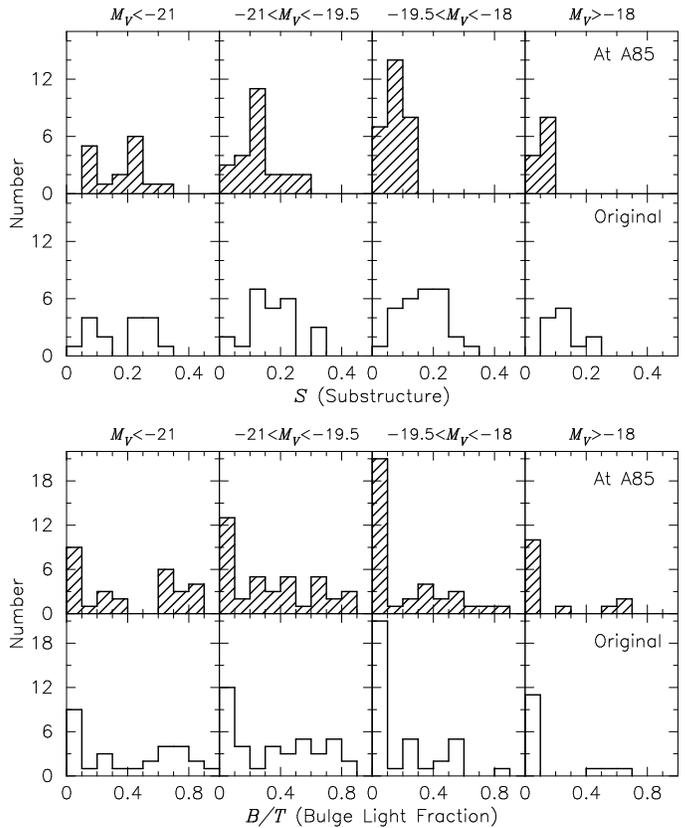}
\caption{
Quantitative morphology distributions in four luminosity bins, from left
to right: $M_V-5\log{h}\leq -21.0$, 
$-21.0<M_V-5\log{h}\leq -19.5$, $-19.5<M_V-5\log{h}\leq -18.0$,
and $M_V-5\log{h}> -18.0$.  The $S$ ({\it top}) and $B/T$ ({\it bottom}) sets of
eight panels are divided into
morphology parameters from fits to the original ({\it outlined bins})
and the redshifted to A85 ({\it hatched bins}) images.  Parameter values
are averages from GIM2D fits to $B$ and $R$-band data.
For comparing substructure we plot only the $S$ distributions
from fits to the IBG and VBG selected field galaxies.
A noticeable change between the original and redshifted morphology 
distributions occur for $S$ at $M_V>-19.5+5\log{h}$.
We compare the full field
sample $B/T$ distributions and find no significant difference between the
original and redshifted $B/T$ distributions in all $M_V$ bins
(K-S test probabilities of being different are $<30\%$, see text).
\label{QMdistr}}
\end{figure}

Following our analysis of the luminosity dependence of $S$ at a
given distance, we construct
similar plots for average ($B$ and $R$ image) $B/T$.
First, we subdivide the $B/T$ distributions into the same four $M_V$ bins
as with $S$ (see bottom portion of Figure~\ref{QMdistr}).  
For the bulge light fraction parameter we consider
all galaxies within the field sample including those with red colors.  
The original and
re-imaged $B/T$ distributions appear quite similar in all brightness bins.
Again we apply a K-S test to compare the rest and shifted distributions
in each luminosity bin and find they are statistically similar with
low probabilities ($<30\%$) of not being drawn from the same parent
sample.

\begin{figure}[hp]
\includegraphics[scale=0.8, angle=0]{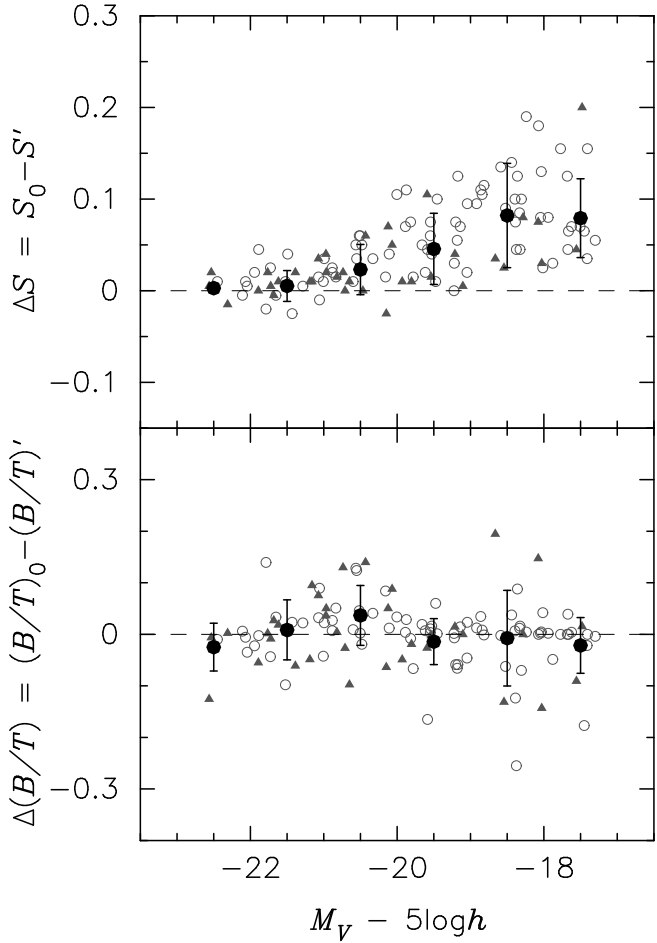}
\caption{
Quantitative morphology differences $\Delta q = q - q^{\prime}$ calculated
for mean ($B$ and $R$ data) parameters from the original 
galaxy appearances ($q$) and from redshifting to A85 ($q^{\prime}$).
The difference in substructure $\Delta S$
({\it top panel}) and bulge light fraction $\Delta B/T$ 
({\it bottom panel}) are 
plotted as a function of intrinsic galaxy luminosity $M_V$.  For each
quantitative morphology difference we distinguish IBG$+$VBG ({\it open
circles}) and RSG ({\it filled triangles}) galaxy types.
The average morphology parameter difference, for $S$ (IBG$+$VBG only) and
$B/T$ (all field galaxies), within $\Delta M_V=1.0$~mag bins from 
$-17.5+5\log{h}$
to $-22.5+5\log{h}$~mag are
given by the large bold solid circles with error bars representing
the $\pm1\sigma$
uncertainties for each mean value.  The mean $\Delta S$ becomes noticeably
nonzero at $M_V\gtrsim-19.5+5\log{h}$; the average $B/T$ difference shows little
deviation from zero for all $M_V<-17+5\log{h}$.
\label{QMdel}}
\end{figure}

In Figure~\ref{QMdel} (bottom) we plot $\Delta B/T = B/T - B/T^{\prime}$
versus $M_V$ for all field sample galaxies.  Although there is
increasing scatter in $\Delta B/T$ towards fainter magnitudes, 
the majority of objects show little change between $B/T$ measurements from
original and redshifted images.

Therefore, we conclude that bulge morphology measurements derived from our
$V$-band cluster observations are reliable for galaxies spanning the full range
of bulge fraction $0<B/T<1$ and total luminosity $-23<M_V-5\log{h}<-17$.
We note that $B/T$ is less affected than $S$
because the effects of pixel binning are incorporated in GIM2D
\citep{im02}, and galaxy bulge light is typically well sampled
({\it i.e.} has much higher S/N than disk).  

%fig5

%fig6

%fig7

\subsection{Estimated Color Profiles}
Our B/D decompositions provide a useful estimate of the internal color
gradient of each galaxy.  In Paper~3
we established that the ratio of
intrinsic half-light radii measured from GIM2D fits to $U$ and $V$-band
images provides a reasonable color gradient estimate 
($CGE \equiv \log{ \slantfrac{r_{\rm hl}(V)}{r_{\rm hl}(U)} }$) for each
cluster member.
Briefly, red color gradients ($CGE<0$) produce
 more peaked $V$-band profiles
($r_{\rm hl}(V)<r_{\rm hl}(U)$) and thus a redder central region, 
while galaxies with
profiles that become bluer towards the center will have $CGE>0$.  If a
galaxy has identical sizes in each passband ($CGE=0$), then we infer that the
$(U-V)$ color distribution across the galaxy is uniform ({\it i.e.} no
color gradient).  The average error in our color gradient estimates is
$\sigma_{CGE}\sim0.045$.

\begin{figure}[hp]
\includegraphics[scale=0.8, angle=0]{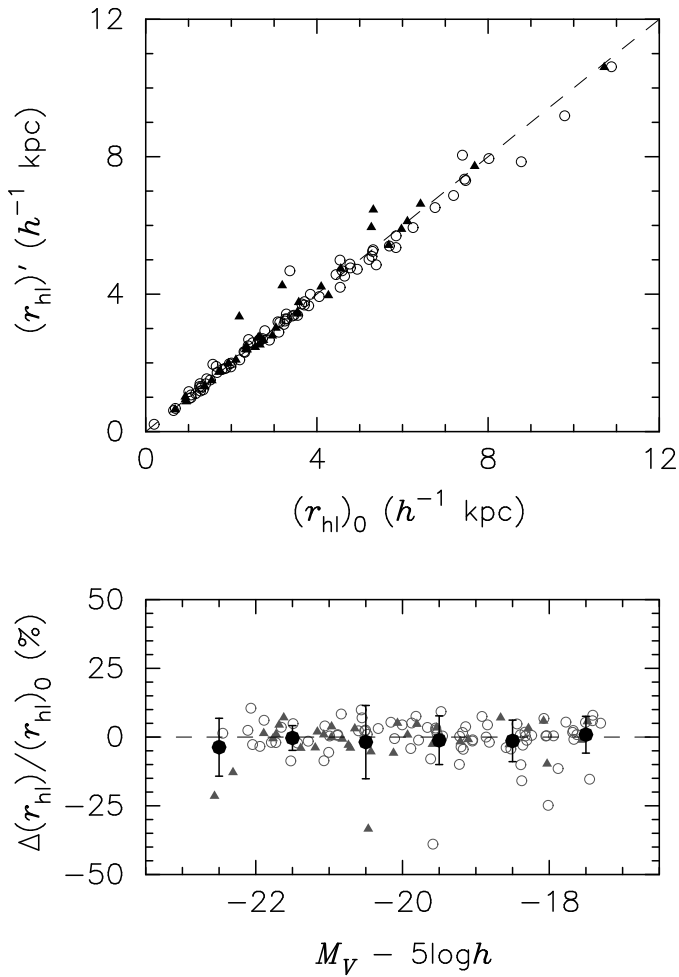}
\caption{
Field galaxy physical sizes from average of GIM2D half-light radii in $B$
and $R$ passbands.  {\it Top panel:} agreement between sizes derived from
original $(r_{\rm hl})_0$ and redshifted to cluster A85 $(r_{\rm hl}^{\prime})$
images.  {\it Bottom:} the percent difference as a function of intrinsic
galaxy luminosity $M_V$.  We distinguish galaxy types: IBG+VBG ({\it open
circles}), RSG ({\it filled triangles}).  The average percent difference
per 1.0~mag bins are given by the large bold solid circles with error 
bars representing the $\pm1\sigma$
uncertainties for each mean value.  There is good agreement between 
$(r_{\rm hl})_0$ and $(r_{\rm hl}^{\prime})$ over all $M_V$.
\label{sizetest}}
\end{figure}

$CGE$ depends directly on our measurement of galaxy size which
may be affected by observational resolution and light
gathering power.  Therefore, we compare the physical sizes measured from
original and artificially redshifted field galaxy images during our tests
of $B/T$ and $S$ reliability.  At a redshift $z$, the physical size is
given by the product of angular size and distance 
$\theta_{\rm hl} \times D_{\rm A}(z)$.  As shown in Figure~\ref{sizetest}, the
average sizes from field galaxy B/D decompositions of 
original $(r_{\rm hl})_0$
and redshifted ($r_{\rm hl}^{\prime}$) $B$ and $R$-band
imaging are in 
excellent agreement over the full luminosity range of our cluster sample at the
distance of cluster A85.
We note that the three galaxies in the bottom panel of Figure~\ref{sizetest}
with percent differences in excess of $25\%$ have poor sky determination due to
proximity to very bright foreground stars.

In Paper~3 we found that more than $50\%$ of blue cluster members
either lack a significant color gradient or have bluer centers relative to
their outer regions.  The color profiles of the blue galaxies were in stark
contrast to the red members which had red gradients in qualitative agreement
with those seen in early-types \citep{peletier90}.

Red color gradients appear to be a common
property of spirals, and they are due to
a combination of age and metallicity effects resulting in the outer disks
having younger ages and lower metallicity measurements \citep{dejong96c}.
Spectroscopic studies have shown that local clusters contain a fair fraction
of galaxies that had a recent episode of enhanced SF 
\citep{couch87,barger96}, or even a starburst 
\citep{caldwell98,poggianti99}.  Furthermore, \citet{bartholomew01}
have found that strong Balmer absorption (so-called ``K$+$A'' galaxies)
cluster members at $z>0.3$ have bluer color gradients indicating
their recent SF activity.
If galaxies entering the dense cluster environment have SF truncation
proceeded by a starburst, especially a centrally located burst 
\citep[as seen in][]{rose01},
their color profiles may be quite different compared to those of normal
field spirals.  Thus, we will compare the blue cluster galaxy $CGE$ results
to similar measures from their field counterparts to look for differences
in centrally concentrated SF histories.

%fig8

%-- 4 COMP----------------------------------------------------------------
\section{Cluster-Field Comparison}
We have analyzed 143 cluster and 78 field galaxies in a 
consistent manner to make direct comparisons between blue galaxy
populations with similar luminosities ($M_V\leq -17+5\log{h}$) that reside in
two distinctly different environments seen in the local Universe.
Specifically, we have performed B/D decompositions on images of cluster
and field blue-selected (IBG and VBG) galaxies.  Recall that VBGs have 
spiral-like colors $\geq0.425$~mag bluer than the well-defined CMR, while
IBGs are intermediately blue in color between the VBG $(U-V)$ cut and red
sequence galaxies within $2 \sigma_{\rm CMR}$ of the CMR.
We have artificially redshifted and reobserved 
the field galaxy data and all A496 ($z=0.033$) galaxies prior to fitting 
to assure that we
compare the structural properties of cluster and field galaxies derived
from similar S/N and resolution quality data -- i.e. at a common distance
corresponding to that of A85 and A754 ($z=0.055$).
In the transformation scenario the field population represents the source
of recent arrivals; therefore, to look for signs of evolution among recent
cluster arrivals we need to make comparisons with their field counterparts.
In this section we compare measures of
bulge light fraction, observed morphological substructure,
estimated color profile gradient, physical size, and luminosity
to quantify whether differences exist between
matched blue galaxy populations in the cluster and field environments.

\subsection{Bulge-to-Total Light Morphology}
Typical $B/T$ values for ellipticals and spheroids are
$B/T>0.5$ \citep[\eg][]{im01,im02}, while later-type
disk-dominated galaxies usually have $B/T<0.3$ \citep[\eg][]{lilly98,im02}.
This coarse separation at $B/T\sim0.3$ into early and late morphologies is
illustrated in Figure~\ref{TvsQM} using a representative field galaxy
sample spanning the full range of Hubble types (E-Sd).  In Paper~3,
we showed that $>75\%$ of red cluster galaxies have $B/T>0.3$, while $>75\%$
of blue members are $B/T<0.3$.  We stress that we are
not making qualitative morphology comparisons and we have not visually 
classified
the cluster members using any standard system.  We will simply make 
cluster-field comparisons using this
quantifiable and repeatable measure of the fraction of light matched by
the best-fit model bulge component.

\begin{figure}[hp]
\includegraphics[scale=0.85, angle=0]{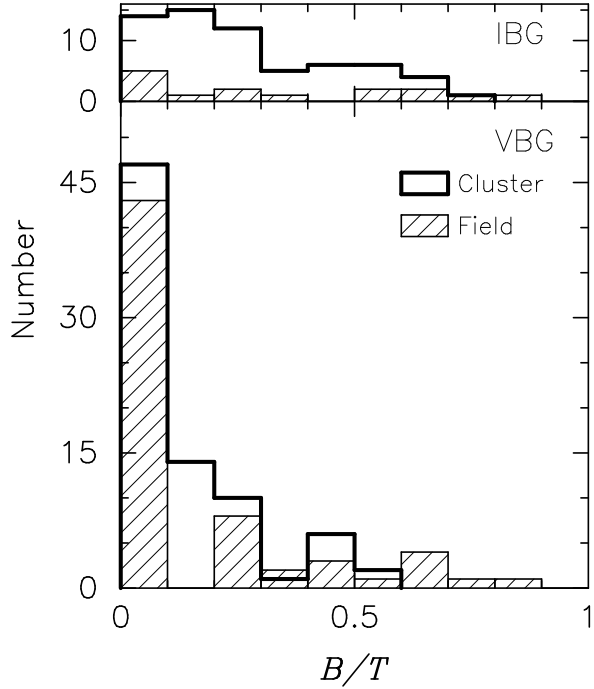}
\caption{
Bulge-to-total light fraction ($B/T$) distributions for IBG ({\it top}) 
and VBG ({\it bottom}) populations.  The cluster ({\it solid outline}) and
field {\it hatched bins}) distributions are similar within each blue galaxy
division (see text).  
The field $B/T$ values
are averaged results from GIM2D fits to the artificially redshifted
$B$ and $R$-band data.  Cluster $B/T$ values are from the higher
S/N $V$-band images.  The cluster and field samples have matched effective
resolution for comparison (i.e. imaged at a common distance of $z=0.055$).
Average errors in $B/T$ range from 25\% for $B/T<0.15$, to 10\% for $B/T>0.30$.
\label{BTdistr}}
\end{figure}

In Figure~\ref{BTdistr}
we plot the $V$-band $B/T$ distributions for the IBG and VBG
populations from the cluster and field environment.  The local field results
are derived from $B$ and $R$-band images 
artificially reobserved at the distance of cluster A85.  All cluster 
measurements are based on imaging at the same distance.
The cluster and field VBGs shown in the bottom panel
have typically low ($B/T<0.3$) bulge fractions,
with the majority peaked at $B/T\leq0.1$.  A K-S test for the cluster 
versus field VBGs shows no significant
difference between the relative fraction of disk light in very blue galaxies
residing in these two disparate environments.  We find a similar result
when comparing the $B/T$ distributions for cluster and field IBGs.
The IBG galaxies (top panel) are 
disk-dominated although they have a much flatter low
$B/T$ ($\leq0.3$) distribution than VBGs.
The cluster IBG and VBG $B/T$ distributions are different 
at $>99.9\%$ confidence, while a similar comparison between the field sample
blue populations shows a moderate difference.  We provide details of the
K-S tests in Table~2.
If we assume that relative color correlates with cluster arrival
time, then the $B/T$ difference between the IBG and VBG populations may
be evidence for processes within the dense cluster that
reduce the disk light profile.  We discuss the physical mechanisms
that might produce evolution from disk-dominated to bulge-dominated
systems in \S5.1.

%fig9

\subsection{Structural Morphology}
\citet{im02} have shown that residual substructure values $S<0.08$ can be used
in conjunction with bulge-dominated $B/T$ measures to successfully
select E/S0 galaxies from a sample of local field galaxies,
without substantial contamination from later-types.  In Figure~\ref{TvsQM}
we note the the majority of E/S0 galaxies from our field sample do indeed
have $S<0.08$, and conversely most spirals (Sa-Sd) show $S\geq0.08$.
Naturally, the degree of SF activity correlates at some level with
the observable amount of spiral substructure.  

\begin{figure}[hp]
\includegraphics[scale=0.85, angle=0]{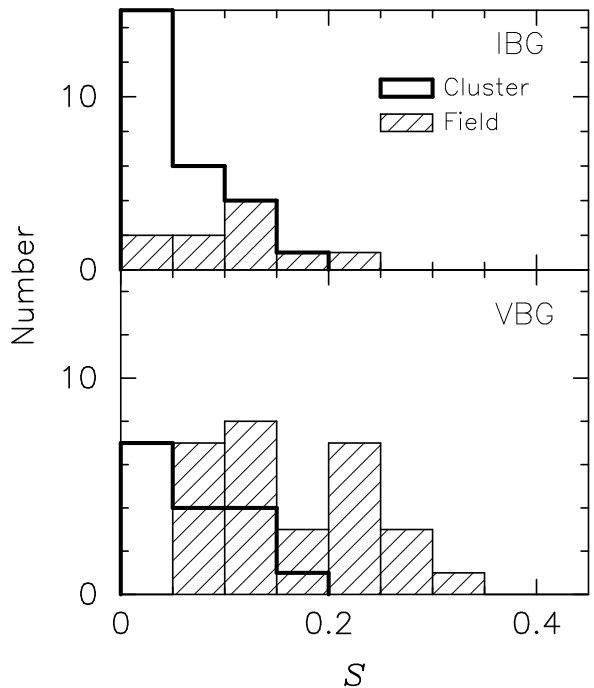}
\caption{
Residual substructure ($S$) distributions for IBG ({\it top}) and
VBG ({\it bottom}) populations brighter than $M_V=-19.5+5\log{h}$.  
Recall, we apply this absolute magnitude cut as the limit for reliable $S$
measurements at the common effective resolution (i.e. imaged at a distance
of $z=0.055$; see \S~\ref{Stest}).  The field $S$ values ({\it hatched bins})
are averaged results from GIM2D fits to the artificially redshifted
$B$ and $R$-band data.  Cluster $S$ values ({\it solid outline}) 
are from the higher S/N $V$-band images.
Cluster VBGs exhibit significantly less substructure than their field 
counterparts (see text).  Typical mean errors in $S$ are $10-20\%$.
\label{Sdistr}}
\end{figure}

In Figure~\ref{Sdistr} we present the field and cluster $V$-band $S$
distributions for IBGs and VBGs brighter than $M_V=-19.5+5\log{h}$.
In \S~\ref{Stest} we used the local field sample to find that we can
reliably measure $S$ for galaxies brighter than this absolute magnitude limit
at the distance of cluster A85 ($z=0.055$).  Recall that all $S$
measurements are from images degraded to the same effective resolution,
and the local field sample results are from the averaged
$B$ and $R$-band NFGS data.
%n Table~2, we give the K-S test results for comparing substructure
%istributions of various samples using three $S$ reliability
%agnitude cuts: $M_V-5\log{h}=$ -20.0, -19.5, and -19.0.
%he findings hold regardless of luminosity cut.  
Recall that we have
removed 27 irregular/peculiar galaxies from our field comparison sample.
None of these galaxies are brighter than the $S$ reliability magnitude
limit, thus, their removal does not affect our results.
%onetheless, we visually inspected the $R$-band images of each
%iscarded irregular/peculiar field galaxy and all exhibit strong 
%tructural features ({\it e.g.}
%opsidedness, multiple nuclei) that, if modeled with a simple B$+$D fit,
%ould produce $S>0.1$ residuals.

As shown in Table~2,
the cluster VBG and IBG substructure morphology distributions given in
Figure~\ref{Sdistr} are not statistically
different from each other.  Using $S>0.1$ to denote galaxies with 
significant residual substructure, most blue cluster galaxies
with $M_V\leq-19.5+5\log{h}$ 
exhibit little substructure.  Roughly $70\%$ of VBGs and
$80\%$ of IBGs have $S\leq0.1$, which is contrary to expectations given
the high fraction of disk-dominated ($B/T\leq0.3$) systems seen in the
previous section.  In the field
environment, we expect blue disk galaxies to have significant substructure
associated with spiral arms and late Hubble types (see Figure~\ref{TvsQM}).
Indeed, in Figure~\ref{Sdistr} we see that $\sim75\%$ of field VBGs have
$S>0.1$, in stark contrast to the clusters.
While the cluster and field VBG populations have similar relative
fractions of disks, there is a
striking difference (at $>99.6\%$ confidence) in their 
residual substructure properties.  Even after diminishing
the observable spiral substructure by re-imaging blue field galaxies
at a distance given by $z=0.055$, they show much stronger deviations from
a smooth light profile compared to cluster members with the same 
effective resolution, color,
luminosity, and $B/T$ morphologies.  The cluster and field IBGs
brighter than $M_V=-19.5+5\log{h}$ are somewhat different in their
substructure distributions.

On average, the cluster VBGs appear smoother, {\it i.e.} have
less substructure ({\it e.g.} spiral arms, \ion{H}{2} regions, bars, SF knots
and dust lanes) than field VBGs.  As discussed by \citet{vandenbergh76}, these 
so-called ``anemic'' spirals have morphological characteristics intermediate
between S0s and normal spirals.
While the quite blue colors of these
galaxies from different environments suggest similar relative fractions
of young stellar populations,
the strong differences in residual substructure suggest different current SF
properties.  Similarly, \citet{balogh98} found a lower mean 
\ion{O}{2}$\lambda3727$ equivalent width in disk-dominated galaxies
near the cluster virial radius compared to the field.  In addition, many
disk-dominated galaxies in X-ray weak clusters have smooth morphologies
and lack emission lines \citep{balogh02b}, which shows a direct connection
between the absence of spiral features and ongoing SF.  
Galaxies in poor groups also exhibit less asymmetric substructure than their
field counterparts \citep{tran01}.  This analysis provides evidence for
morphological smoothing, presumably related to SF activity.  It is not clear
whether the smoothing takes place in groups or subclusters prior to reaching
the cluster, upon entry to the cluster,
or sometime thereafter.  We contemplate the possible origins of structural
morphology
transformation of the infalling disk galaxy population 
in the discussion section.

%fig10

\subsection{Estimated Color Gradients}
In Paper~3 we showed that $CGE$ represents a reasonable estimate of
a galaxy's profile color gradient using GIM2D half-light radii from fits to
red and blue passband images.  We found that most blue cluster galaxies, 
especially VBG types, lacked the significant red color gradient exhibited by
the bulk of the red members.  Our focus here is to determine whether 
blue galaxies in the field and clusters show differences in their overall 
color profiles.  For our field galaxy comparison sample we set
$CGE=\log{ \slantfrac{r_{\rm hl}(R)}{r_{\rm hl}(B)} }$ and use
half-light radii derived from the
redshifted $B$ and $R$-band images.  This is a crude measure of
color gradient, thus, subtle differences due to different choices of blue
and red passband should have little consequence when trying to discern
if a galaxy is redder or bluer towards its center.

\begin{figure}[hp]
\includegraphics[scale=0.85, angle=0]{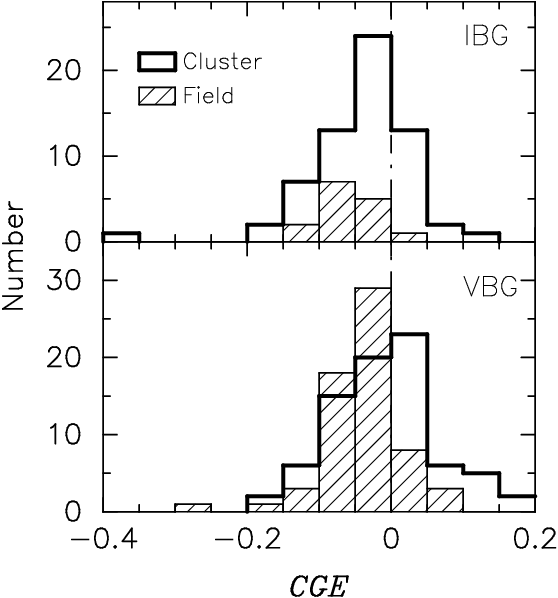}
\caption{
Color gradient estimates $CGE$ for IBG ({\it top}) and VBG ({\it bottom})
types.  $CGE$ is given by the ratio of
half-light radii measured from fits to red and blue passband images:  
$V$ and $U$-band for the cluster data ({\it solid outline});
$R$ and $B$-band for the field data ({\it hatched bins}).
The cluster and field samples have matched effective
resolution for comparison (i.e. imaged at a common distance of $z=0.055$).
Contrary to field VBGs with typically red centers
($CGE<0$), a large number of cluster VBGs have profiles with overall
blue colors ($CGE\sim0$) or blue cores ($CGE>0$).  We include a {\it dot-dashed
line} showing $CGE=0$.  Typical mean errors are $\sigma_{CGE}\sim0.045$.
\label{CGEdistr}}
\end{figure}

We plot the $CGE$ distributions for IBGs and VBGs from
cluster and field environments in
Figure~\ref{CGEdistr}.  The blue field galaxies have $CGE$ distributions that
are redder on average than their counterparts found in clusters.  The field
results are in qualitative
agreement with red color gradients observed in
late-types \citep[\eg][]{dejong96c}.  A K-S test shows
a significant ($>99.9\%$) difference between the
cluster and field VBG $CGE$ distributions (see Table~2) given in 
the bottom panel.  We find the fraction of VBGs with
estimated blue color gradients ($CGE>0$) to be $46\%$ in clusters and 
only $18\%$
in the field.  A similarly strong difference in relative fractions of blue
gradient IBGs exists between the cluster ($24\%$) and field ($0\%$)
samples (top panel), although the overall cluster and field IBG distributions
are not statistically different.  This result implies that, unlike
normal star-forming field spirals, many of the bluer cluster members
are blue throughout or even show blue cores.  This appears to be true
especially for cluster VBGs.

Color profile differences are presumably evidence for a recent
deviation from ``normal'' SF activity.  Overall blue colors may suggest
an episode of fairly strong global SF within the past few gigayears, while 
blue centers imply recent or on-going centrally concentrated SF.  For
example, these uniformly blue cluster galaxies are consistent
with populations of galaxies found in distant ($z>0.3$) clusters
showing enhanced Balmer absorption spectra \citep{couch87,barger96}.
Additionally, \citet{bartholomew01} found that distant cluster galaxies with
strong Balmer absorption have blue color gradients, and are presumed
to be a post-starburst population \citep{caldwell98,poggianti99}.
Furthermore,
\citet{rose01} found that currently star-forming galaxies in local
clusters often show blue cores and have more centrally concentrated emission
lines than field galaxies.  
%here are many theoretical mechanisms that
%ay trigger centralized SF; e.g. global tidal fields \citep{byrd90}, or
%am-pressure stripping under special circumstances 
%citep{fujita99,abadi99,vollmer01}, to name but a few.
As with the structural morphology differences, the color gradient difference 
we find between very blue cluster and 
field members provides further evidence that the VBG populations in
these two environments have undergone quite different recent star formation
histories.  We discuss physical processes that may produce blue color
gradients in cluster galaxies in the discussion section.
We note that recent starburst
activity could correlate with increased luminosity in the bluer cluster 
galaxies; however, we find no preference for brighter IBG and VBG
members to have $CGE>0$.

Certainly, our finding that many recent cluster arrivals lack typical red
color gradients should be investigated further.  This result is based on
simple global size measurements in different passbands.  A formal color
gradient analysis of this large cluster galaxy sample will be addressed
in a future paper.  Additionally, we will include a study of fiber
spectra from the central regions ($1.5$~\hkpc\ at cluster A85) of each
cluster member.

%fig11

\subsection{Scaling Relations}
Galaxy sizes and how they scale with brightness are fundamental observables.
Our 2D B/D decompositions of cluster galaxy $V$-band light profiles
yield PSF de-convolved, 
and thus intrinsic, half-light radii \rhl\ and
disk scale lengths $h_0$.  In the following comparison, we estimate
$V$-band field galaxy
sizes using the average radii from GIM2D B/D
decompositions of NFGS $B$ and $R$-band imaging redshifted to the distance
of cluster A85.

\subsubsection{Global Scaling Relations}
We have established that red and blue cluster members occupy somewhat
different regions of luminosity-size ($L-r$) space
(Paper~3).  Briefly, cluster galaxies with colors redder than VBGs show
a broad correlation between absolute magnitude and size, similar to that 
seen for local ellipticals, while VBG members are
concentrated around $M_V=-19+5\log{h}$ and $r_{\rm hl,circ}=1$~\hkpc.  Moreover,
cluster VBGs appear to concentrate towards lower mean surface brightness
when compared to fiducial values from large field spiral studies
\citep[\eg][]{burstein97}.
In the previous sections we have
observed differences between structural properties of blue galaxies
selected from clusters and the field; therefore, it is likely that scaling
relations may be affected by environment as well.

First we compare the global $L-r$ relations 
found in blue selected cluster
and field systems.  Our $B/T$ analysis shows that these galaxies typically 
contain significant disk components, therefore, we correct the
absolute magnitudes for internal extinction and 
the global half-light sizes to face-on values.
The circular half-light radius is 
$r_{\rm hl,circ} = r_{\rm hl,sma} \sqrt{b/a}$, where 
$r_{\rm hl,sma}\equiv r_{\rm hl}$ is the half-light semi-major axis
given by GIM2D, and $b/a$ is the overall inverse
galaxy axis ratio that we estimate from the disk inclination ($b/a=\cos{i}$) 
for $B/T<0.5$ systems, or from the bulge ellipticity
($b/a=1-\epsilon$) for $B/T\geq0.5$ galaxies.  The
internal extinction corrected absolute magnitude is
$M_V^i=M_V-2.5C_{\rm abs}\log(a/b)$.
The amount of internal disk absorption is governed by
the factor $C_{\rm abs}$ which varies from one (optically thick) to
zero (optically thin).  How $C_{\rm abs}$ is computed is highly debated in the
literature \citep[see][]{giovanelli94,tully98}.  For our $V$-band 
measurements we use $C_{\rm abs}=0.53$ which is extrapolated from internal
extinction corrections given in \citet{burstein95}.  Both the cluster and
field magnitudes have been corrected for Galactic extinction using
\citet{schlegel98} and \citet{burstein82}, respectively.  A slight
$\lesssim0.05$ magnitude difference may exist between the two samples
because of the different dust corrections used.

\begin{figure*}
\includegraphics[scale=0.9, angle=0]{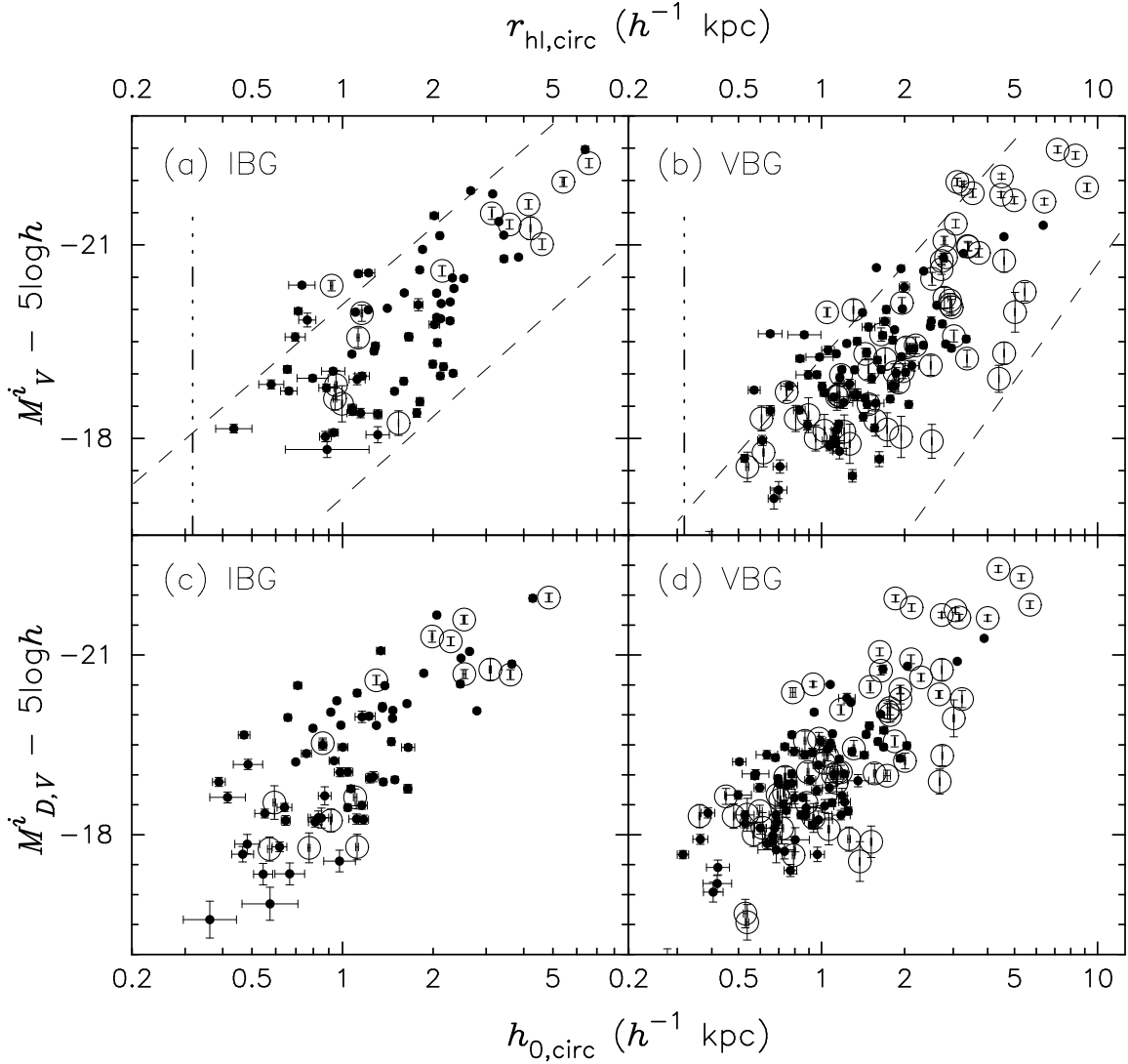}
\caption{
Global and disk luminosity versus size correlations for blue-selected
galaxies.  All parameters are $V$-band corrected to face-on values
(see text for details).
{\it Top}: (a) IBG and (b) VBG half-light 
radius against absolute magnitude; {\it bottom}: (a) IBG and (b) VBG
disk scale length against absolute disk luminosity.
Cluster galaxies are given as {\it small filled
points}, while {\it large open circles} depict field members.
The {\it sloped dash lines} in the top two panels represent the region of 
face-on $L-r$ space populated by local (a) Sa-Sbc and 
(b) Sc-Sdm galaxies from \citet{burstein97} adapted by \citet{simard99},
which we translate to $V$-band assuming mean galaxy colors given in 
\citet{fukugita95}.  The {\it vertical dot-dash lines} in (a) and (b) show
the physical resolution limit.
}
\label{BLUEscale}
\end{figure*}

For the cluster and field samples split into two blue C-M types, we plot
half-light radius against $V$-band absolute magnitude in the top two
panels of Figure~\ref{BLUEscale}. 
We see that the cluster and field IBGs
follow similar relations between
face-on size and inclination-corrected luminosity.  For
comparison, we delineate the region of $L-r$ parameter space occupied by local
spirals corrected for internal absorption \citep{burstein97,simard99},
and transformed to $V$-band quantities 
using average colors of $(B-V)=0.65$ for Sab/Sbc types \citep{fukugita95}.
The VBG populations (Fig.~\ref{BLUEscale}b)
have luminosities and sizes that fall within the region defined by local
Sc-Sdm spirals from \citet{burstein97,simard99}, using $(B-V)=0.50$.
Yet it is clear that,
unlike the field, clusters contain very few
large ($>2$~\hkpc), bright ($<-20+5\log{h}$) VBG members.  We are confident in the
spectroscopic and photometric completeness of the cluster data at 
$M_V<-18.2+5\log{h}$.  In \S4.4.3 we establish
that the bright end of the field VBG luminosity function
studied here is fairly representative.

%fig12

\subsubsection{Disk Scaling Relations}
Besides the global scaling relations seen in our cluster galaxies, we
analyze similar correlations for the disk components of the IBG and VBG
populations.  Using GIM2D derived disk parameters we compute the internal
extinction corrected disk luminosity $M_{D,V}^i = M_V^i - 2.5\log(1-B/T)$,
and the face-on disk scale length $h_{0\rm ,circ} = h_0\sqrt{\cos{i}}$, where 
$h_0$ is the semi-major axis scale length.  We plot the disk component
$L-r$ relations for blue galaxies in the bottom two panels of
Figure~\ref{BLUEscale}.

The disks of blue cluster members occupy the same regions of
disk luminosities and scale lengths as
their field counterparts.  The disk correlations have similar slopes and
scatter as their global analogies shown in Figures~\ref{BLUEscale}a,b.
Cluster IBG disks (panel c) and all blue field galaxy 
disks (panels c and d) are evenly distributed over
the observed correlation from small and faint 
(0.5~\hkpc, $-17+5\log{h}$) to large and bright
(5.0~\hkpc, $-22+5\log{h}$).  Yet most of the more recent cluster arrivals
(panel d) have disks that are even
more tightly concentrated towards small sizes and low luminosities than when
shown using global scaling relations (panel b).  The paucity of bright
($M_{D,V}^i<-20+5\log{h}$), large 
($h_{0,{\rm circ}}>1.5$~\hkpc) disks among the VBG
cluster members, compared to field spirals with equally blue colors,
is likely related to the difference we measure between the morphological
substructure of VBGs in these two environments -- i.e. galaxies with
small, faint disks will exhibit less substructure.  

\begin{figure*}
\includegraphics[scale=0.9, angle=0]{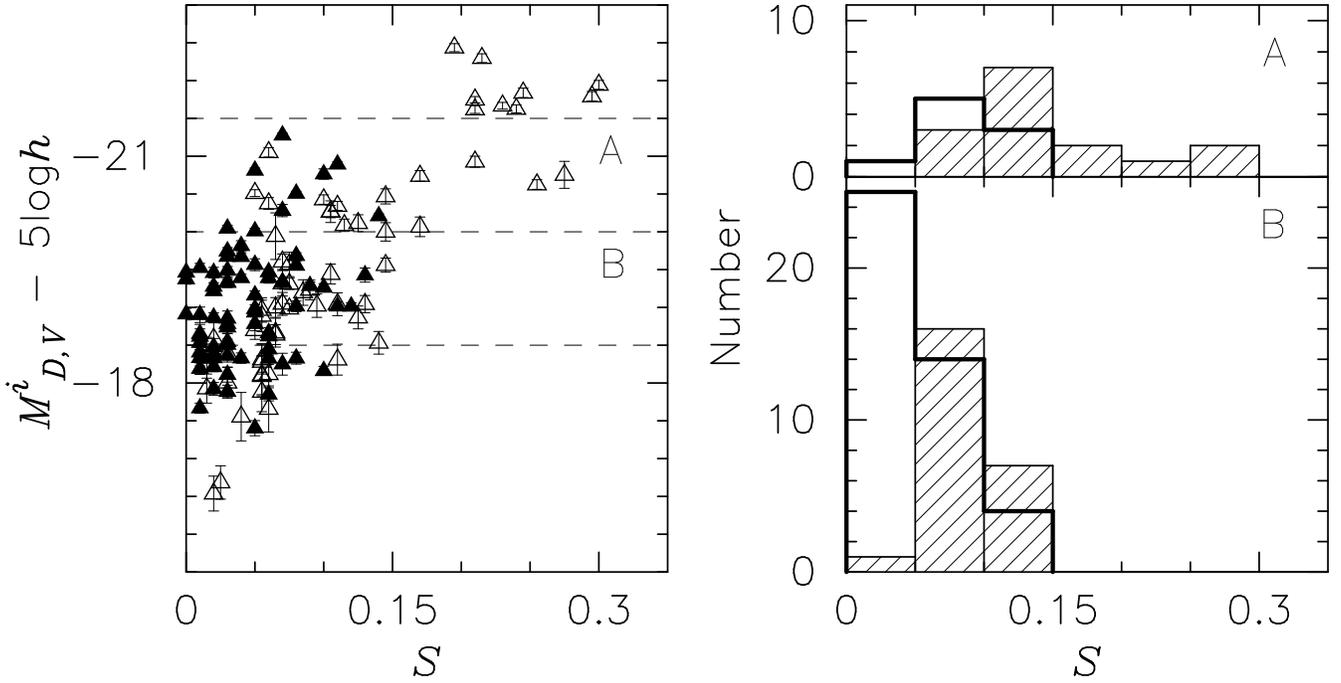}
\caption{
Structural morphology as a function of disk luminosity for VBGs in cluster and
field environments ({\it left panel}).  
The disk luminosity is $V$-band corrected for inclination.
We distinguish between field galaxies ({\it open triangles}) and
cluster members ({\it filled triangles}).  We delineate two (A and B)
disk luminosity bins with {\it dashed lines}.
The cluster and field $S$ distributions for disk luminosity bins A and B
are given in the {\it right panel}.  Cluster
blue galaxies with lower $S$ are given by {\it solid outline} and their field
counterparts are show as {\it hatched bins}.
}
\label{diskstr}
\end{figure*}

Up to now we have shown
that, in general, cluster and field
VBGs spanning a similar range in luminosity have
quite different $S$ distributions.  Here, we test whether the lack of
substructure in cluster VBGs contains an environmental component
independent of luminosity, or whether the $S$ distributions differ solely
because of the differences 
we find between the blue galaxy luminosity functions of our cluster and 
field samples.  In Figure~\ref{diskstr} we plot $S$ as a function 
of $M_{D,V}^i$ for cluster and field VBGs.  We see that substructure and disk
luminosity correlate in VBG populations as expected; nevertheless, the spread 
along $S$ in this relation
appears governed by the global environment.  Within any disk luminosity bin,
the division between low substructure (cluster) and high substructure (field)
remains among blue selected galaxies.  We illustrate this for two disk
luminosity bins in the right panel of Figure~\ref{diskstr}.
We surmise that a physical process such as galaxy harassment
\citep{moore96b,moore98,moore99} is responsible
for decreasing disk size, luminosity, and substructure in
recent cluster arrivals.  We note that harassment is most effective
on lower mass objects such as the blue cluster galaxies we study
here.  We discuss galaxy harassment and other possible mechanisms 
for transforming
the appearance of the latest cluster members in \S5.1.

%fig13

\subsubsection{Bright End Luminosity Functions of Very Blue Galaxies}
In our global and disk scaling relation analyses we notice
the bright end ($M_V<-20+5\log{h}$) of the
cluster and field VBG luminosity functions are quite different.  In
Figure~\ref{BLUEscale}b we find very
few bright VBGs in local clusters compared to field spirals with equally blue 
colors.  This luminosity difference appears even more pronounced 
in Figure~\ref{BLUEscale}d when only the disk
components are considered.  Our finding is
interesting if the distributions of bright blue galaxies in the two
environments reflect the local Universe.  

At $M_V\leq-18.2+5\log{h}$ the cluster sample is $\sim85\%$ complete 
spectroscopically, thus, we are
confident that the relative number of cluster galaxies brighter than this limit
are representative of rich clusters out to roughly a virial radius.
To rule out field sample
selection biases, we show the luminosity function of our field VBGs in
Figure~\ref{nfgsLfcn}.  The bold outline shows the overall fraction
of galaxies per absolute $B$-band magnitude bin for the full NFGS sample.
\citet{jansen00} have selected these objects to sample the varying 
morphological mix as a function of brightness, and to approximate the local
galaxy luminosity function of \citet{marzke94}.  Next we show the
$M_B$ distributions for all NFGS galaxies that obey our VBG selection
criteria (dash-dot outline), and the subsample of 63 field VBGs that
we select (see \S \ref{fieldsel}) for our analysis (binned histogram).
We apply a K-S test to compare the luminosity distributions of the two field 
VBG samples at the bright end.  For $M_B<-18$ 
(corresponding to $M_V\lesssim -19+5\log{h}$)
we find a $<1\%$ probability that the
subsample of VBGs that we select are different from the total set of 
\citet{jansen00} galaxies in the VBG region of C-M space.
Therefore, if the NFGS provides a representative sampling of the local
field population, we assume that
the bright VBGs in our field sample are fairly representative also.

We plot the luminosity distributions of the cluster and field VBGs in
Figure~\ref{vbgLF}.  Given the above arguments, we assume that our
cluster and field selections of VBGs brighter than $M_V=-19+5\log{h}$
are characteristic of their corresponding environments.
We see a striking difference
in the bright end luminosity functions of very blue selected
galaxies in high and low density regions of the local Universe.  
Specifically, VBGs brighter than $L^{\star}$ are much more abundant in
the field than in clusters.  We note that if we include the bright VBGs in the
NFGS that we exclude in our field sample selection, the observed difference 
between cluster and field LFs for bright blue galaxies would be even larger.
It is well-known that the high luminosity population of         
rich clusters is largely early-types
(red galaxies), while the field is spiral-rich \citep{oemler74}.
Previous studies of the environmental
dependence of galaxy luminosity functions have shown an overall excess of bright
galaxies in clusters relative to the field \citep{goto02}.
Here we demonstrate that
the membership of rich clusters is deficient of bright, blue galaxies
in contrast with the field.

%fig14
\begin{figure}[hp]
\includegraphics[scale=0.6, angle=0]{mcintosh_fig14.ps}
\caption{
Fraction of galaxies per total absolute $B$-band magnitude for the NFGS
\citep{jansen00}.  The total NFGS sample selection (198
galaxies, {\it bold outline}) approximates the local galaxy luminosity 
function of \citet{marzke94}.  Furthermore, we show the
distributions of all NFGS galaxies meeting our VBG C-M based definition
({\it dash-dot outline}), and the
subsample of VBGs that we analyze here ({\it binned histogram}).
A K-S test (see text) demonstrates
that our field VBG sample selection (see \S \ref{fieldsel}) 
is a fair representation of
all NFGS VBGs brighter than $M_B=-18$,
corresponding to $M_V\lesssim -19+5\log{h}$.
}
\label{nfgsLfcn}
\end{figure}

%fig15
\begin{figure}[hp]
\includegraphics[scale=0.85, angle=0]{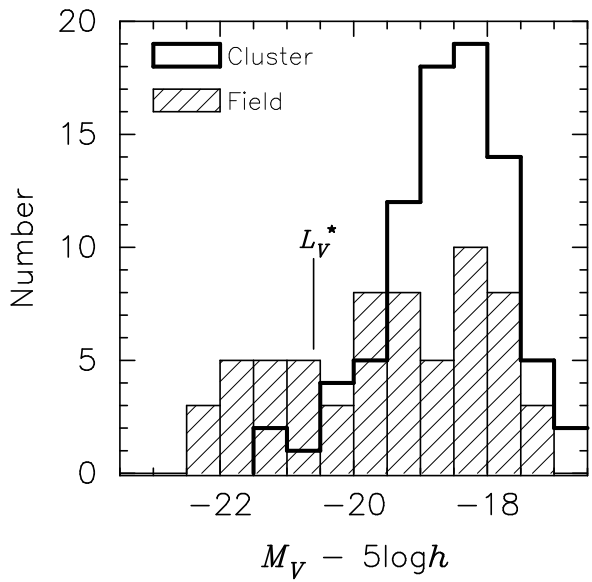}
\caption{
Comparison of the VBG $V$-band luminosity distributions in 
the cluster ({\it solid outline}, $N=80$ galaxies) and field 
({\it hatched bins}, $N=63$ galaxies) samples.
At brighter than $M_V\sim -19+5\log{h}$ the cluster and field VBG selections are
fair representations of their environments.
$L^{\star}$ (at $M_V = -20.6 + 5\log{h}$) is shown for reference.
}
\label{vbgLF}
\end{figure}

%-- 5 DISCUSS------------------------------------------------------------
\section{Results and Discussion}
In this paper we have compared blue
galaxies residing in clusters, to galaxies of similar color in field 
environments,
through a detailed analysis of their $V$-band 
structural properties from 2D profile B/D decompositions.  Our cluster
sample consists of 143 blue galaxies drawn from the combined catalog of
spectroscopically confirmed members in local clusters: A85 ($z=0.055$),
A496 ($z=0.033$) and A754 ($z=0.055$).  The field comparison sample of
78 blue galaxies comes from \citet{jansen00}.  Both samples are limited
to $M_V\leq -17+5\log{h}$.  We have degraded and
resampled the field galaxies and the cluster A496 members to measure 
their structural properties at a matched effective 
resolution to clusters A85 and A754 at $z=0.055$.
We have divided each of the two samples into very blue
galaxy (VBG) and intermediately blue galaxy (IBG) populations (see Table~1),
based on the relative color difference $\Delta(U-V)$ between the
$(U-V)$ color of each galaxy and the well-defined CMR.
Evolutionary synthesis
models following the C-M evolution of field galaxies that suffer
SF truncation upon entering the overall cluster environment, with or without 
an initial starburst, confirm that relative
blueness tells us something about cluster accretion time and ``membership age''
\citep[\eg][]{bicker02}.  In particular, we are interested
in the relative properties of the most recent cluster arrivals (VBGs) and
field counterparts of the same color and luminosity -- ``normal'' spirals 
thought to be the supply of infalling galaxies
in an hierarchical framework.  The observed difference of
cluster galaxies is thought to be driven by the morphological transformation
of newer members.  In each cluster we find significant fractions
($18-23\%$) of VBG and IBG galaxies that have spatial and kinematic properties
consistent with being a recently arrived population
(see Paper~2 for details).  Here we discuss the results of our blue galaxy
cluster-field comparison in relation to the transformation scenario.

Our findings are summarized as follows:
\begin{enumerate}
\item The VBG population in clusters has typically low bulge light
fractions ($B/T\sim0.1$), consistent with disk morphologies.  The
$B/T$ distribution of cluster VBGs 
is well matched to that of field spirals, while
the cluster IBGs show a broader $B/T$ distribution that is significantly
different ($>99.9\%$ confidence) than the bluer cluster members,
with a ratio of 3/4 disk-dominated ($B/T<0.3$) to 1/4 bulge-dominated 
($B/T>0.3$) systems.

\item The distribution of residual substructure morphology differs
between cluster and field
VBGs at $99.6\%$ confidence.  Even after diminishing
the observable spiral substructure in field VBGs by artificially redshifting
to the maximum cluster redshift, these GIM2D fitted galaxies have much 
stronger deviations from
a smooth light profile in contrast to cluster members with the same color,
luminosity, and $B/T$ morphologies.  Even when considering the
brightest ($M_V<-20+5\log{h}$) VBGs, cluster disks show little
evidence of spiral arm substructure associated with current SF.  The cluster
IBGs show an average lack of substructure similar to the more recent arrival
VBG population. 

\item From our estimates of color gradient we find that cluster VBGs have 
color profiles that lack any clear inward reddening ($CGE<-0.1$).
Half of the recent
cluster arrivals are globally blue or even have blue cores 
($CGE\geq 0$).  The IBG cluster population
likewise has a fair fraction of blue cores.  Moreover, the estimated color
profiles seen in cluster VBGs are quite different ($>99.9\%$ confidence)
from those exhibited in normal field spirals.

\item For both the blue cluster and field populations, 
selected by C-M properties
and analyzed in the same manner, the $L-r$ relation for either the total
galaxy or just its disk
falls in the same range of parameter space defined by field
surveys, if we identify IBGs with Sa-Sbc types and VBGs with Sc-Sdm.
However, at the bright end of the VBG luminosity functions,
where we are confident in
the completeness for both cluster and field membership, the cluster
population has a striking deficiency of bright ($M_V<-20+5\log{h}$) VBGs
compared to the field.  In contrast to field VBGs, the cluster VBG population
tends to concentrate towards having small ($<2$~\hkpc), 
faint ($M_V>-20+5\log{h}$)
disks.  Furthermore, the VBG disk properties appear linked to the lack
of significant morphological substructure seen in these presumed recent cluster
arrivals.
\end{enumerate}

To summarize, the relative differences we find between very blue cluster 
members at large projected radii, and field galaxies with matched colors
and luminosities analyzed in the same manner, rule out the dense
cluster core as the {\it only} site where cluster galaxy transformation 
occurs.  This result is qualitatively similar to the work of \citet{gomez03}
who found that the SFRs of late-type cluster galaxies are diminished,
at cluster-centric distances (3-4 virial radii) well outside of the 
cluster core, compared with the field population.
We note that our data are not sufficient to
favor a single well-defined transformation process.  
For example, we lack observations
at large enough cluster-centric radii to extend to the cluster-field
transition region where transformation may begin.  Moreover, we 
are magnitude-limited and thus unable
to explore the cluster dwarf galaxy population, which could harbor 
transformed galaxies in some scenarios \citep[e.g.][]{moore99}.
Nevertheless, in what follows we interpret our findings by examining 
the predictions
of several popular transformation mechanisms.

\subsection{Galaxy Transformation Mechanisms}
Several physical processes have been suggested to explain the
transformation of infalling, star-forming spirals into red cluster S0s.  
Major mergers are known to create ellipticals \citep{toomre72}, and may represent
the origin of the homogeneous population of {\it older} cluster ellipticals
\citep{kauffmann96}; however, this mechanism is likely unimportant
considering the high relative velocities ($\sigma \sim 1000$~\kms)
found in massive present-epoch clusters \citep{vandokkum99}, 
and the steadily declining merger rate to the
present-day \citep{dressler94a,couch98}.  Furthermore, only in special
cases can mergers produce S0s \citep{bekki01,cretton01}.
The removal of hot gas, followed by
the subsequent consumption of available fuel with no replenishing source,
may be responsible for
disk-dominated galaxies with no signs of ongoing SF
\citep[``strangulation'' or ``starvation'',][]{larson80,balogh00,balogh02b,bekki02}.  For example,
halo gas stripping can gradually transform normal star-forming spirals
into red cluster S0s with a passive (or anemic) spiral as a
transitionary phase \citep{bekki02}, yet this process has trouble 
explaining low disk substructure with
simultaneous blue colors as seen in cluster VBGs.

The process of galaxy harassment \citep{moore98,moore99}, acting on
infalling galaxies within the cluster virial radius, works well to
explain the bulk of our observations under the following assumptions:
(1) accreted galaxies follow infall orbits like those presented in
\citet{bekki02}, such that galaxies take several gigayears to move from
cluster-centric distances of 2 Mpc to 1 Mpc; (2) the transformation
is ongoing and slow (due to the nature of the orbit); 
and (3) the progenitors of cluster VBGs were late-type spirals
(i.e. Sc-Sm), while IBGs were earlier types (Sa-Sb).
%subsubsection{Galaxy Harassment}
Galaxy harassment is the tidal induced evolution of a disk galaxy from
multiple, high-speed
interactions with massive galaxies and the cluster gravitational field
\citep[originally][]{moore96}.
In simulations harassment is effective at transforming
late-type disks (cluster VBG progenitors) into dwarf ellipticals (dEs)
or dwarf spheroidals \citep{moore96b,moore98,moore99}, and
early-type spirals (cluster IBG progenitors) into S0s \citep{moore99}.
These transformations occur primarily through disk heating and stripping
from multiple encounters with other cluster members.  The harassment
timescale is $\sim1-3$ gigayears, with the effectiveness greatest towards the 
cluster center where interactions are more frequent \citep{sensui99};
therefore, galaxies on more radially extended orbits may take longer to 
accumulate harassment effects due to the lower number density of galaxies 
at larger cluster radii.

Galaxy harassment provides a reasonable process for explaining
the relative differences between cluster and field VBGs.  Harassment
is most effective on lower mass (luminosity) galaxies such as the VBGs
we study in this paper.
The lack of bright ($M_V,D<-20+5\log{h}$),
large ($h_0>2$\hkpc) VBG disks in local clusters is consistent
with the expectations of harassment where the large scale length ($>7$\hkpc)
disks of low surface brightness galaxies are decreased to $h_0<2$\hkpc\ 
through severe ($50-90\%$) stripping of their stellar disks \citep{moore99}.
We show in Figure~\ref{diskstr} that decreased substructure in VBGs is 
correlated with disk luminosity; therefore, 
the significant contrast between the morphological substructure
of cluster and field VBGs is highly suggestive of harassment which claims
to remove spiral substructure through disk heating \citep{moore99}.  
Finally, the cluster VBGs show evidence of recent SF activity in their
color profiles, with some fraction exhibiting blue cores
consistent with the large post-starburst population seen in
clusters \citep{caldwell98,poggianti99,bartholomew01}.
Harassment may initiate central starbursts, especially in lower luminosity
spirals (Sc-Sm), by setting up instabilities
that funnel gas towards the galaxy center \citep{fujita98,abadi99}.

The three conditions we impose above constrain the effects of harassment
to be consistent with the spiral-like colors and disk-dominated morphologies
of cluster VBGs.  Specifically, assumptions (1) and (2) mean that not
enough time, nor enough interactions, have transpired to fully strip away 
the disk of a late-type progenitor (condition 3).  Moreover,
even though the cluster VBG disks are smaller, they maintain a small
$B/T$ measure if their progenitors were nearly pure disks to begin with.
In other words, stripping a pure large disk merely produces a small disk.

In addition, the observed properties of cluster IBGs are consistent with
the predictions of galaxy harassment if condition (3) holds -- namely
the progenitors of cluster IBGs were early-type spirals.  For this case,
the effects of galaxy harassment on these high surface brightness (HSB)
systems is less severe than for low surface brightness later types
\citep{moore99}.  Early types will transform into S0s, that is
bulge-dominated galaxies with smooth appearances, in agreement with cluster 
IBGs that have
significantly different $B/T$ distributions compared to the disk-dominated
VBGs, yet have equally little substructure.  Furthermore, the cluster
IBGs have a higher fraction of larger ($h_0>1.5$~\hkpc) disks
compared to the VBGs, which is consistent with galaxy harassment because
this process will heat, but not destroy, the disks of HSB galaxies
\citep{moore99}.  In fact, many of these moderately blue cluster members are
likely the modern-day counterparts to the blue S0s found in more distant 
($z>0.3$) clusters \citep{vandokkum98,rakos00,smail01}.
We note that some cluster IBGs, especially the fraction that are small
and faint, may be further evolved galaxies along a sequence from VBG to IBG.
As such, the progenitors of these galaxies 
would be the same as VBGs, but accreted at earlier
times and thus midway to a dE morphology, the cluster VBG endpoint
in the galaxy harassment scenario.

Finally, it is likely that gas stripping is responsible ultimately for
reducing SFRs in accreted galaxies and producing predominantly red
cluster populations over time, especially as infalling 
galaxies enter the dense central
regions \citep{valluri90,abadi99,quilis00,bekki02}.
%subsubsection{Ram-Pressure Stripping}
Ram-pressure stripping, first proposed by \citet{gunn72}, is the removal
of the cold, neutral gas reservoir in a star-forming spiral by its rapid
motion through the hot ICM.  Stripping appears responsible for the
\ion{H}{1} deficiencies in cluster later types
\citep[\eg][]{magri88,vollmer01}, and for X-ray wakes trailing behind
ellipticals observed in nearby clusters \citep{drake00}.  Recent 3D simulations
show this process is efficient at quickly ($\sim10^7$~yrs)
removing gas from disk galaxies and thus
suppressing SF \citep{abadi99}.
Adding the effects of turbulent and viscous stripping, \citet{quilis00}
find that $100\%$ of the \ion{H}{1} gas in a luminous spiral like the
Milky Way is removed within $\sim10^8$~yrs
via this mechanism when the galaxy passes through a dense cluster
core ($R_{\rm core} \sim 250$~\hkpc).  These authors find that
stripping quickly extinguishes SF and, thus,
provides an explanation for the enhanced Balmer absorption observed in many
cluster galaxies \citep[\eg][]{couch87,barger96}.
Yet the required close core passage presents a
problem with invoking ram-pressure stripping to explain the properties we
observe in the blue cluster galaxy population.  This process is likely
limited to work well only within the cluster core
\citep{abadi99,quilis00}; therefore, stripping is probably unimportant
for the IBG and VBG members that are not yet physically located near the
dense cluster center (Paper~2).  Nevertheless, the blue and moderately 
blue cluster galaxies must eventually experience stripping when their
orbits bring them within $R_{\rm core}$.  

We note that given the spiral-like colors
of cluster VBGs, it appears that morphological transformation is
separate from, and begins before, color evolution.  This intriguing results
runs contrary to previous work on more distant clusters that find
red galaxies with late-type morphologies (passive spirals) suggesting
color evolution precedes morphological transformation
\citep[e.g.][]{couch98,poggianti99,dressler99}.  We note that we find
examples of rare passive spirals in our cluster sample (Paper~2), which
underscores the complex nature of galaxy transformation and the
likelihood that multiple processes must be important at different times
and under varying conditions.
For example, there exists observational evidence that
ram-pressure stripping is occurring well outside of cluster cores
\citep{neumann01}.  In addition, galaxies on semi-circular
orbits that pass within a few $R_{\rm core}$ may experience a slow gas
removal resulting in a SF decline rather than a truncation \citep{kodama01a}.
This is similar to the predicted effects of hot halo gas stripping
\citep[e.g.][]{larson80,balogh00,balogh02b,bekki02}.
Nevertheless, while it is possible that hot or cold gas stripping can explain
certain unique examples found within the cluster sample,
it cannot explain simultaneously the very blue
colors, blue cores, and lack of morphological substructure we find in
the bulk of cluster VBGs. 

%-- 6 CONCL.------------------------------------------------------------
\section{Conclusions}
The morphological transformation of field (or filament) 
spirals during infall into the
dense cluster environment \citep[\eg][]{kodama01b} has been proposed to explain 
the rapid evolution of cluster galaxies over the past 5~gigayears.
In particular, the spiral-to-S0 ratio in clusters has decreased from $\sim2:1$
at $z=0.5$ to $\sim1:2$ at present day, suggesting a direct evolutionary link
between the decreasing numbers of spirals and increasing numbers of
S0s \citep{dressler97}.  In this paper we uncover evidence for 
environment-driven
galaxy transformation through a detailed comparison of recent cluster
arrivals with galaxies of similar luminosities and blue colors in
the field.  In a Universe {\it without environmental dependent evolution
outside of the dense cluster cores}, we would expect blue disk galaxies
inhabiting field and cluster regions to have similar
morphology, size, and color gradient distributions.
Our findings show conclusively that fundamental galaxy
properties do indeed reflect the environment in which the galaxy is found.  

We find structural differences between the blue galaxies inhabiting nearby
($z<0.06$) clusters, compared to field environments.  The
majority of blue cluster members are physically smaller and fainter than
their field counterparts.  At a matched size and luminosity, the
newer cluster arrivals have quantifiably less internal substructure, yet have
equally disk-dominated morphologies as normal field spirals.
Furthermore, half of blue cluster galaxies have blue cores or
globally blue color profiles in contrast with field
spirals which typically show redward color gradients.
Blue cores suggest enhanced nuclear star formation, possibly a starburst, while
uniformly blue profiles are consistent with an episode of fairly strong
global star formation in the past few gigayears.  We show in Paper~2 that
blue cluster galaxies are members of a recently infalling population that
has not yet encountered the cluster core.  Therefore, the differences we
observe between very blue cluster and field galaxies show that galaxy
transformation occurs in accreted galaxies {\it before} the violent
effects (e.g. strong tides, ram-pressure) of the cluster core come
into play.

Much of what we observe in the blue galaxy populations of local clusters can be
explained by the process of galaxy harassment
\citep{moore96,moore96b,moore98,moore99}, under several imposed conditions.
Nevertheless, ram-pressure stripping
\citep{gunn72,abadi99,quilis00} must play an important role 
in assuring that SF is eventually quenched in cluster galaxies to
produce the strong color evolution, especially towards the cluster
core, implied by the Butcher-Oemler effect
\citep[][and references therein]{bo84,rakos95,margoniner01}.
The fact that transformation has occurred in cluster galaxies with
spiral-like colors (the VBGs) shows that the processes that govern color
(SF) and physical morphology evolution are decoupled.

\acknowledgments{}
We thank the anonymous referee for much effort and care in the reviewing
process, which improved this paper substantially.
We are grateful to Luc Simard for much hands-on support using GIM2D.
Rolf Jansen is thanked for making the NFGS images available.
We acknowledge helpful discussions and correspondence with Eric Bell,
Roelof de Jong, Neal Katz, Rob Kennicutt, Ben Moore, Matthias Steinmetz,
Ann Zabludoff and Dennis Zaritsky.
Finally DHM thanks the remainder of his thesis committee -- Craig Foltz, Chris
Impey, Ed Olszewski, and Daniel Eisenstein -- for insightful comments that
ultimately improved this paper.
This research has made extensive use of NASA's Astrophysical Data
System Abstract Service (ADS) and the astro-ph/ preprint server.
We acknowledge support from NSF grant xxxx.

\appendix
\section{Artificially Redshifting and Reobserving the Field Data}
To ``move'' a galaxy out to a further distance,
we must artificially degrade the observed
image to mimic a larger redshift and different seeing conditions by
reducing the flux and smearing the effective resolution.  Then we rebin
the flux and add noise to construct 
a new image as if observed with the telescope and detector used to obtain the
cluster data.  

For each field galaxy we first determine the amount of flux dimming 
$\Delta_i^{\rm dim}$ by requiring that the rest-frame $M_V$ magnitude
(estimated in \S3) is conserved for the observed and artificially redshifted 
images.  Recall that we are using both the NFGS $B$ and
$R$-band images and transforming these intensity values to rest-frame
$V$ to match our cluster observations configuration.
The expected $V$-band flux of each galaxy that is artificially redshifted
to $z^{\prime}=0.055$ and reobserved through our 0.9-meter/Mosaic system is then
\begin{equation}
f^{\prime} = 10^{-0.4(M_V + DM^{\prime} - k_V^{\prime} - zp_{\rm sys})} ,
\end{equation}
where $DM^{\prime}=36.18-5\log{h}$ is the cosmological distance modulus, 
$k_V^{\prime}=-0.098$~mag is the $V$-band \citet{poggianti97} $k$-correction,
and $zp_{\rm sys}=26.90$ refers to the total system zero point from the $V$
cluster observations (for a 600 second exposure at 1.33 airmass).  
We reduce the flux of each galaxy image by the ratio
\begin{equation}
\Delta_i^{\rm dim} = \frac{f^{\prime}}{f_{\rm tot}(z_i)} ,
\label{DimEqn}
\end{equation}
where $f_{\rm tot}$ is the total model flux from the B/D decomposition
to the original NFGS image observed at redshift $z_i$.

Next, we want each re-imaged galaxy to have the same effective resolution
(0.76~\hkpc) as our furthest cluster imaging.  To achieve this seeing we
convolve the original, dimmed image with a Gaussian of width
\begin{equation}
\sigma = \frac{\sqrt{({\rm FWHM}^{\prime})^2 - ({\rm FWHM}_i)^2}}{2.354} ,
\end{equation}
where ${\rm FWHM}_i$ is the image's original full-width at half-maximum
seeing and ${\rm FWHM}^{\prime}$ is the desired resolution given by
\begin{equation}
{\rm FWHM}^{\prime} = \frac{D_{\rm A}(z^{\prime})}{D_{\rm A}(z_i)} \cdot \frac{1.02\arcsec}{p_i} .
\end{equation}
The target seeing disk is related to the ratio of physical sizes of an
object at redshift $z^{\prime}$ and at $z_i$,
which are given by their corresponding angular diameter distance $D_{\rm A}(z)$
(=153~\hmpc for A85).
The additional factors are the $V$-band mode seeing ($1.02\arcsec$) in the
A85 image and the original NFGS image pixel scale $p_i$.  Following 
\citet{hogg00}, $D_{\rm A}(z) = D_{\rm C}(z) / (1+z)$ where $D_{\rm C}(z)$ is
the line-of-sight comoving distance for a given redshift $z$.  Assuming
a flat $\Omega_k=0$ Universe,
\begin{equation}
D_{\rm C}(z) = \frac{c}{H_0} \int_0^z \left[ \Omega_{\rm M}(1+\bar{z})^3 + \Omega_{\rm \Lambda} \right]^{-1/2} d\bar{z} .
\label{ComovdEqn}
\end{equation}

Finally, to reimage each galaxy as if through the telescope and detector used
during observations of cluster A85, we rebin each galaxy image by a factor 
\begin{equation}
\Delta_i^{\rm rebin} = \frac{D_{\rm A}(z_i)}{D_{\rm A}(z^{\prime})} \cdot \frac{p_i}{p^{\prime}} ,
\end{equation}
where $p^{\prime}=0.423\arcsec {\rm pix}^{-1}$ is the Mosaic pixel scale.
The additional factor of pixel scale ratio is necessary
because we are rebinning to a {\it different} pixel scale
than the original images.  We rebin each image with flux conservation.
To reproduce the noise characteristics of our cluster imaging, we add the A85
mean $V$ sky level with random Poisson noise using the effective Mosaic
gain.  We apply the full procedure to each $B$ and $R$ image from our
field galaxy sample.

%--BIBLIOGRAPHY----------------------------------------------------------------

%--TABLES----------------------------------------------------------------------
%% Deluxetables are one-column floats which need to be placed at end of doc.
%% when emulateapj format.

\onecolumn

%====================== table1
\begin{deluxetable}{lcccccccc}
\tablewidth{0pt}
\tablenum{1}
%tabletypesize{\tiny}
\tabletypesize{\small}
\tablecolumns{9}
\tablecaption{Cluster Properties}
\tablehead{\colhead{Cluster} & \colhead{$\alpha_{2000}$} & \colhead{$\delta_{2000}$} & \colhead{$<cz>_{\rm clus}$} & \colhead{$\sigma_{\rm clus}$} & \colhead{$N_{\rm mem}$} & \colhead{$N_{\rm IBG}$} & \colhead{$N_{\rm VBG}$} & \colhead{$R_{\rm vir}$} \\
\colhead{(1)} & \colhead{(2)} & \colhead{(3)} & \colhead{(4)} & \colhead{(5)} & \colhead{(6)} & \colhead{(7)} & \colhead{(8)} & \colhead{(9)} }
\startdata
A85  & 00 41 50.5 & -09 18 11.6 & 16607 & 993 & 180 & 17 & 33 & 1.94\\
A496 & 04 33 37.8 & -13 15 43.5 & 9910  & 728 & 146 &  7 & 27 & 1.37\\
A754 & 09 08 32.4 & -09 37 46.5 & 16369 & 953 & 311 & 39 & 24 & 1.32\\
\enddata
\tablecomments{Abell cluster name (1) and coordinates (2,3) of the central cD 
galaxy.  The mean cluster recessional velocity (4) and internal
velocity dispersion (5), in units of km~$s^{-1}$, are given from the
spectroscopic survey of \citet{christlein03}.
Total number of spectroscopically
confirmed members (5), number of IBGs (6), and number of VBGs (7).
In (8) we give the virial 
radius (\hmpc) from data compiled in \citet{girardi98}.}
\end{deluxetable}
%====================== table1

%**********table2
\begin{deluxetable}{llcccccc}
\tablewidth{0pt}
\tablenum{2}
%tabletypesize{\tiny}
\tabletypesize{\small}
\tablecolumns{8}
\tablecaption{Cluster and Field K-S Test Results}
\tablehead{\colhead{} & \colhead{} & \multicolumn{2}{c}{$B/T$} & \multicolumn{2}{c}{$S$} & \multicolumn{2}{c}{$CGE$}\\
\cline{3-8} \\
\colhead{Sample 1} & \colhead{Sample 2} & \colhead{$N_1:N_2$} & \colhead{$(\%)_{\rm diff}$} & \colhead{$N_1:N_2$} & \colhead{$(\%)_{\rm diff}$} & \colhead{$N_1:N_2$} & \colhead{$(\%)_{\rm diff}$} \\
\colhead{(1)} & \colhead{(2)} & \colhead{(3)} & \colhead{(4)} & \colhead{(5)} & \colhead{(6)} & \colhead{(7)} & \colhead{(8)}
}
\startdata
cluster VBG & field VBG   & 80:63  & 44.3    & 16:29  & 99.6 & 80:63  & $>99.9$ \\
cluster IBG & field IBG   & 63:15  & 72.1    & 26:10  & 92.9 & 63:15  & 83.6 \\
cluster VBG & cluster IBG & 80:63  & $>99.9$ & 16:26  & 1.6  & 80:63  & 96.4 \\
field VBG   & field IBG   & 63:15  & 94.4    & 29:10  & 83.3 & 63:15  & 83.6 \\
\enddata
\tablecomments{Descriptions of the pair of samples we use for each K-S test are 
in (1) and (2).  The number of galaxies in the two samples and the K-S
probability that the distributions are not drawn from the same parent sample
are given for $B/T$ (3 and 4), for $S$ (5 and 6), and for $CGE$ (7 and 8).
The number of galaxies we use for $S$ comparisons are reduced
due to the $M_V\leq-19.0+5\log{h}$ reliability limit for $S$ measurements
at the common effective resolution.}
\end{deluxetable}

\end{document}